\documentclass[12pt,preprint,apj]{emulateapj}

\newcommand{\kms}{{km s$^{-1}$}~}
\newcommand{\kmsMpc}{{km s$^{-1}$ Mpc$^{-1}$}~}

\usepackage{epstopdf}
\usepackage{color}

\shorttitle{Abell 2744}
\shortauthors{Lee \& Jang}

\begin{document}

\title{
Globular Clusters, Ultracompact Dwarfs, and
Dwarf Galaxies in Abell 2744 at the Redshift of 0.308 
}
\author{Myung Gyoon Lee  \& In Sung Jang}
\affil{Astronomy Program, Department of Physics and Astronomy, Seoul National University, Gwanak-gu, Seoul 151-742, Korea}
\email{ mglee@astro.snu.ac.kr, isjang@astro.snu.ac.kr}


\begin{abstract}
We report a photometric study of globular clusters (GCs), ultracompact dwarfs (UCDs), and dwarf galaxies in the giant merging galaxy cluster Abell 2744 at $z=0.308$.  
Color-magnitude diagrams of the point sources derived from deep F814W (restframe $r'$) and  F105W (restframe $I$) images of Abell 2744 in the Hubble Space Telescope Frontier Field
show a rich population of point sources whose colors are similar to those of typical GCs. 
These sources are as bright as  $-14.9<M_{r'}\leq -11.4$ ($26.0<$F814W{\rm (Vega)}$\leq 29.5$) mag, being mostly UCDs and bright GCs in Abell 2744.
The luminosity function (LF) of these sources shows a break at $M_{r'}\approx -12.9$ (F814W $\approx28.0$) mag, indicating a boundary between UCDs and bright GCs.
The numbers of GCs and UCDs are estimated to be
$385,000\pm24,000$, and $147\pm26$, respectively.
The clustercentric radial number density profiles of the UCDs 
and  bright GCs show similar slopes, but these profiles 
are much steeper than that of the dwarf galaxies and  the mass density profile based on  gravitational lensing analysis. 
We derive an LF  of the red sequence galaxies for $-22.9<M_{r'} \leq -13.9$ mag. 
The  faint end of this LF is fit well by a flat power law with $\alpha=-1.14\pm0.08$, showing no faint upturn. 
These results support the galaxy-origin scenario for bright UCDs: they are the nuclei of dwarf galaxies that were stripped when they pass close to the center of massive galaxies or a galaxy cluster, while some of the faint UCDs are the bright end of the GCs. 
\end{abstract}


\section{INTRODUCTION}

The brightest galaxies in galaxy clusters host rich globular cluster (GC) systems as well as satellite galaxies:
e.g., M49 and M87 in Virgo, NGC 4874 in Coma, NGC 6166 in Abell 2199, and the brightest galaxies in Abell 1689 \citep{gei96,lee93,har09, pen11, ala13, har14, har16}. 
Galaxy clusters host not only the GCs around galaxies, but also intracluster GCs between galaxies. 
These intracluster GCs are mostly metal-poor and they are found even far from massive galaxies \citep{lee10,pen11,ala13,dur14}. 

The nearest galaxy clusters like Virgo and Fornax are excellent targets to investigate the details of stellar populations and galaxy populations. However, these galaxy clusters occupy a wide field in the sky so  it is not easy to cover the entire area of a galaxy cluster.
On the other hand, distant galaxy clusters provide an advantage that it is possible to cover the entire area of a galaxy cluster and its background field. 
Thus they can serve as 
a very useful target to study global properties of stellar populations and galaxy populations in an entire cluster, if deep images with high resolution are available.

To date Abell 1689 at $z=0.1832$  is the most distant galaxy cluster ($d=861$ Mpc and $(m-M)_0=39.68$ 
for $H_0=73$ \kmsMpc) in which GCs were studied in detail.
\citet{ala13} selected point sources with $27.0<$F814W({\rm AB})$<29.32$ mag ($26.4<$F814W({\rm Vega})$<28.9$ mag, corresponding to $-13.3 <M_r <-10.8 $ mag) in the deep Hubble Space Telescope (HST)/Wide Field Camera (WFC) F814W images, and estimated the total number of GCs.
They obtained a value for the GCs at $R< 400$ kpc from the cluster center, $N_{GC}=162,850^{+75,450}_{-51,310}$, showing that Abell 1689 hosts the largest GC system among the known galaxies or galaxy clusters.
They also found that the radial mass density profile of the GCs at $R<400$ kpc is steeper than those of the total mass and X-ray gas, while it is flatter than that of the stellar light.
However, no color information is available for these point sources, because they used only one-band (F814W) images for their study.

\begin{deluxetable*}{ccccccccc}
\tablecolumns{9}
\tablewidth{0pc}
\tablecaption{Basic Properties of Abell 2744}
\tablehead{
\colhead{Parameter} & \colhead{Value} & \colhead{References} }
\startdata
Heliocentric Velocity & 92336 \kms & 1,2  \\
Redshift & $z=0.308$ & 1,2 \\
Luminosity Distance$^a$ & $1540\pm90$ Mpc & 1 \\
Distance Modulus$^a$ & $(m-M)_0  = 40.94\pm0.15$ & 1 \\
Scale$^a$ & 4.370 kpc arcsec$^{-1}$ = 262.18 kpc arcmin$^{-1}$ & 1 \\
Age at Redshift$^a$  &   9.937 Gyr & 1 \\
Foreground Extinction & $A_V=0.036$, $A_I=0.020$,  $A_J=0.009$ & 3 \\
Velocity Dispersion & $\sigma_v = 1,497\pm57$ \kms & 4 \\ 
Mass  & $M(R<200 {\rm kpc} = 45\arcsec {\rm (core)})=2\times 10^{14} M_\odot$& 7 (gravitational lensing) \\
 & $M(R<1 {\rm Mpc})=1.4-2.4\times 10^{15} M_\odot$& 5,6 (galaxy dynamics) \\
Cluster Type  & BM III & 1 \\
X-ray Luminosity & $2.10 \times 10^{45}$ erg s$^{-1}$ & 8 \\ 
\enddata
\label{tab_a2744basic}
\tablenotetext{a}{NED values for $H_0 = 73$ \kmsMpc, $\Omega_M=0.27$, $\Omega_\Lambda = 0.73$.}
\tablecomments{References: (1) the NASA/IPAC Extragalactic Database (NED); (2) \citet{str99};
(3) \citet{sch11}; (4) \citet{owe11}; (5) \citet{bos06}; (6) \citet{mer11}; (7) \citet{jau15}; (8) \citet{gov01}.}
\end{deluxetable*}

On the other hand, \citet{mie04} studied bright point sources in the much shallower ACS F475W (SDSS $g'$), F625W ($r'$), F775W ($i'$), and F850L ($z'$) images of Abell 1689, finding more than 10 ultracompact dwarfs (UCDs) with $i'<26.8$ mag ($M_V < -12.6$ mag).  They selected point sources with stellarity values $>0.6$ derived using  SExtractor \citep{ber96}. They  suggested that the cumulative radial distribution of the UCDs with $i'<26$ mag ($M_V < -13.4$ mag) is consistent with that of the genuine dwarf galaxies in the same cluster, but the size of the sample they used is too small to be statistically significant.

In this study, we present the results of our survey of GCs and UCDs in Abell 2744 (AC 118) at the redshift of $z=0.308$,
which is much more distant than Abell 1689. 
We adopt the cosmological parameters: $H_0 = 73\pm5$ \kmsMpc, $\Omega_M=0.27$, and $\Omega_\Lambda = 0.73$, as in the NASA Extragalactic Database (NED). For these parameters, the luminosity distance to Abell 2744 is  $1540\pm90$ Mpc (corresponding to $(m-M)_0  = 40.94\pm0.15$), and the time elapsed since the Big Bang for the cluster redshift is 9.9 Gyr. 
At this distance, one arcsec (one arcmin) corresponds to
4.37 kpc (262 kpc) so the GCs and the UCDs in Abell 2744 are not resolved in the HST images. 
The basic properties of Abell 2744 are listed in 
{\bf Table \ref{tab_a2744basic}}.

Abell 2744  is a giant galaxy cluster, and is often called the Pandora cluster
because it is known to be in the phase of active major merging \citep{cou84,cou87,owe11}.
Abell 2744 shows several distinguishable features.  
Gravitational lensing analysis (strong lensing and weak lensing), X-ray emission, and galaxy kinematics of Abell 2744 all show
it is one of the most massive galaxy clusters, having a mass of $\sim$2 $\times 10^{15} M_\odot$ \citep{mer11,ric14,wan15,jau15,med16}. 
The mass of Abell 2744 is comparable to that
of the Bullet cluster at a similar redshift \citep{par12}.
The center of the mass distribution in Abell 2744 is close to the center of the southern core where the two brightest cluster galaxies (BCGs) are located. 
It is embedded in a giant radio halo, and are surrounded by huge radio relics \citep{gov01,orr07}.
It has a high X-ray luminosity, but the peak of the X-ray emission is $\sim$250 kpc ($0\arcmin.7$) offset in the north-west direction from the center of the mass map based on the gravitational lensing analysis (see Figure 1 in \citet{mer11}).

Abell 2744 shows a large velocity dispersion ($\sigma_v = 1500 - 1800$ \kms) and two subcomponents that have different kinematics with a mean radial velocity difference of 3000--4000 \kms\citep{bos06,owe11}.
The color-magnitude diagram (CMD) of this cluster shows a population of faint blue galaxies in addition to a red sequence, which is known as the Butcher-Oemler effect \citep{cou84,cou87}. 
Abell 2744 is known to have a high star formation rate, ${\rm SFR_{UV+IR}}=201\pm9 M_\odot$ yr$^{-1}$ for $R<1.1$ Mpc,  and the fraction of IR-detected star-forming galaxies in this cluster is 0.08, comparable to those of other relaxed clusters at similar redshifts  \citep{raw14}.
Four jellyfish galaxies were found in this cluster,
showing evidence of a major cluster merger transforming rapidly the morphology of galaxies. \citep{owe12}.
Abell 2744 hosts no cD galaxies so  it is classified as Bautz-Morgan Type III. Two BCGs are located in the southern core.  These BCGs are No.1 and 2 in \citet{cou84}(their finding chart is given in \citet{cou87}), called CN-1 and CN-2 hereafter. A mass density map derived from gravitational lensing analysis shows peaks at these two brightest galaxies \citep{lam14,wan15,jau15}, and the highest peak is close to the position of the BCG in the south (CN-2). 
Abell 2744 hosts numerous dwarf galaxies as well as massive galaxies, so it is an excellent target for the study of the galaxy luminosity function as well.
Recently, \citet{bla15} reported a detection of GCs and compact star clusters in the single band (F814W) images of Abell 2744.

In this study we investigate the GCs and UCDs as well as dwarf galaxies in Abell 2744 detected in the deep HST Frontier Field (HFF) images obtained with F814W and F105W filters.
Primary goals of observing several massive galaxy clusters in the HFF are to study the gravitationally-lensed sources and to detect and study high redshift sources behind the galaxy clusters. However, these images are also an excellent resource to study the objects belonging to the target galaxy clusters. 
\citet{mon14} studied the properties of the diffuse intracluster light in Abell 2744 using F435W, F814W, F105W and F140W images  in the HFF program.

We select and study the GCs, UCDs, and dwarf galaxies in Abell 2744, taking advantage of the high spatial resolution and depth of these images.
The wide coverage of a large fraction of Abell 2744 in the HFF data provides a unique opportunity of being able to 
investigate any relation among the GCs, UCDs, and dwarf galaxies in a massive galaxy cluster.
This paper is composed as follows.
Section 2 describes data reduction.
In \S 3 we present the CMDs 
of the point sources and extended sources in Abell 2744.
Detected point sources are mostly bright GCs and UCDs in Abell 2744, while extended sources are galaxies.
We investigate the spatial distributions of the GCs and UCDs and  of the galaxies.
We then present  the luminosity functions (LFs) of the GCs and UCDs and of the red sequence galaxies in Abell 2744.
Then we discuss primary results.
Finally we summarize our main results.

\begin{deluxetable}{lcccccccc}
\tablecolumns{9}
\tablewidth{0pc}
\tablecaption{A Summary of HST Data of Abell 2744, Parallel Field, and HXDF}
\tablehead{
\colhead{ } & \colhead{Abell 2744} & \colhead{Parallel Field} & \colhead{HXDF}}
\startdata
Right ascension		& 00 14 21.26	& 00 13 53.35	& 3 32 38.82   \\
Declination			& --30 23 47.3	& --30 22 52.7	& --27 47 28.3 \\
Galactic longitude  & 8 50 08.59	& 9 08 51.13	& 223 33 15.26 \\
Galactic latitude   & --81 14 39.3	& --81 09 17.0	& --54 23 43.6  \\
$A_{F814W}$	& 0.020			& 0.019			& 0.012 \\
$A_{F105W}$	& 0.013			& 0.012			& 0.008 \\
Exposure, F814W& 104,270s 		& 107,766s		& 50,800s \\
Exposure, F105W& 68,952s		& 67,330s		& 266,700s \\
Pixel scale, F814W	& $0\farcs03$ & $0\farcs03$ & $0\farcs03^a$ or $0\farcs06^b$ \\
Pixel scale, F105W	& $0\farcs03$ & $0\farcs03$ & $0\farcs06$ \\
Drizzle software 		& DrizzlePac  & DrizzlePac  & APSIS  \\
\enddata
\label{tab_data}
\tablenotetext{a}{Used for  DAOPHOT photometry.}
\tablenotetext{b}{Used for  SExtractor photometry.}
\end{deluxetable}

\section{DATA AND DATA REDUCTION}

\subsection{Data}

We used the data for Abell 2744 
released as part of the HFF 
program (ID13495, PI:J. Lots, ID13386, PI:S.Rodney, ID11689, PI:R. Dupke) \citep{lot14,koe14}. We selected 
ACS/F814W($I$) and WFC3/F105W($Y$) images among the HFF data set, because we believed this combination of images to be the most efficient and  suitable for searching for GC-like objects in Abell 2744.
The HFF data cover two HST fields in Abell 2744:
one for the southern core region of Abell 2744 and one for the parallel field at $5\farcm7$ west.
Each of these fields covers $\sim 1\farcm96 \times 2\farcm18$.
Basic information of the HST data for these fields is summarized in {\bf Table \ref{tab_data}}.

\begin{figure}
\centering
\includegraphics[scale=0.8]{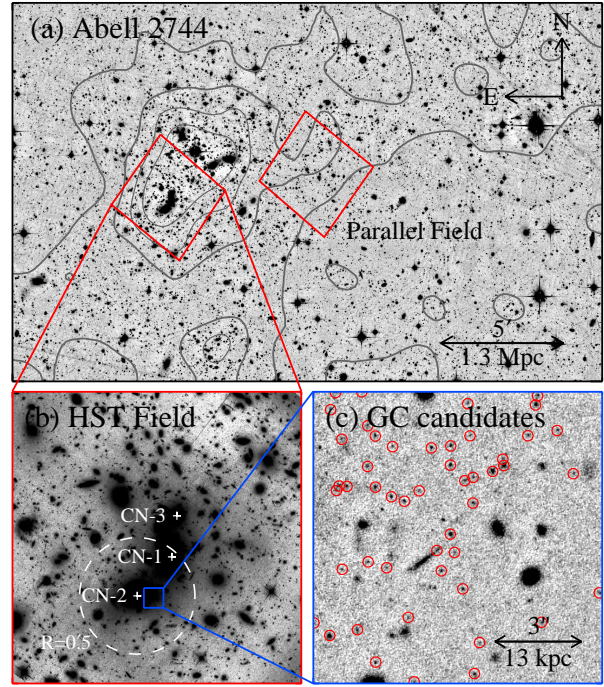} 
\caption{(a) A finding chart for the Abell 2744 HST field (left box) and its parallel field (right box), based on the gray scale map of a CFHT/MegaCam $i$-band image. Contours represent the relative number density of galaxies detected in the MegaCam image. North is up, and East to the left. 
(b) A gray scale map of the HST/ACS F814W image of the Abell 2744 field. 
Crosses denote CN-3, CN-1, and CN-2 from top to bottom. 
(c) A zoom-in image of a small section of the F814W image close to the brightest galaxy, CN-2. Smooth background light has been subtracted from the original image. Numerous point sources (red circles) are seen, which are mostly candidates for GCs and UCDs in Abell 2744.
}
\label{fig_finder}
\end{figure}

{\bf Figure \ref{fig_finder}(a)} displays a finding chart for the HST Abell 2744 field and its parallel field based on the CFHT/MegaCam $i$-band  image of Abell 2744 in the CFHT archive. We overlaid the number density contours of galaxies we derived from the MegaCam image in the same figure.
The Abell 2744 field is located in the highest number density region  of galaxies in the cluster. 
On the other hand, the parallel field is in the region with a number density that is much lower than that of the cluster region, but is slightly higher than that of the background field region. 

\begin{figure}
\centering
\includegraphics[scale=0.9]{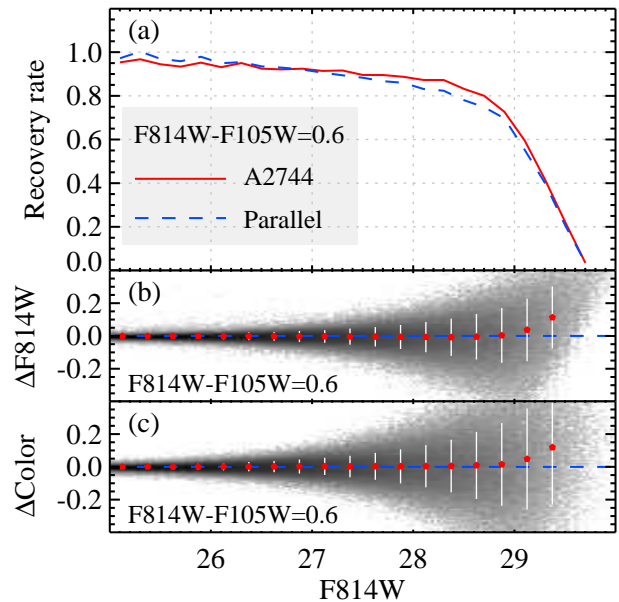} 
\caption{Results of artificial star experiments.
(a) Completeness (recovery rate) vs. F814W magnitudes of input artificial stars with (F814W--F105W)$=0.6$  in the Abell 2744 field (solid line) and the Parallel field (dashed line). 
(b) F814W magnitude differences of input and output artificial stars. 
(c) (F814W--F105W) color differences of input and output artificial stars.
Red circles with errorbars in (b) and (c) represent mean values. 
}
\label{fig_completeness}
\end{figure}

 The effective wavelengths of the F814W and F105W filters for the redshift ($z=0.308$) of Abell 2744 (6220 \AA~  and 8030 
\AA) correspond approximately to SDSS $r'$ and Cousins $I$ (or SDSS $i'$) in the rest-frame,  respectively. 
The total exposure time is $104,270s$ for F814W and
$68,952s$ for F105W.
We used the drizzled science images with a pixel scale of $0\farcs03$ in the released data. 
One pixel corresponds to $4370 \times 0.03 = 131$ pc at the distance to Abell 2744. 
The full width at half-maximum (FWHM) values of the point sources in the images are  $2.9 \sim 3.0$ pixels
($0\farcs087$ to $0\farcs090$, and 380 to 393 pc).
It is expected that GCs ($r_{\rm eff} <10$ pc) and UCDs ($10< r_{\rm eff} <100$ pc)  at the distance to Abell 2744 appear as point sources in these images.

\begin{figure*}
\centering
\includegraphics[scale=0.8]{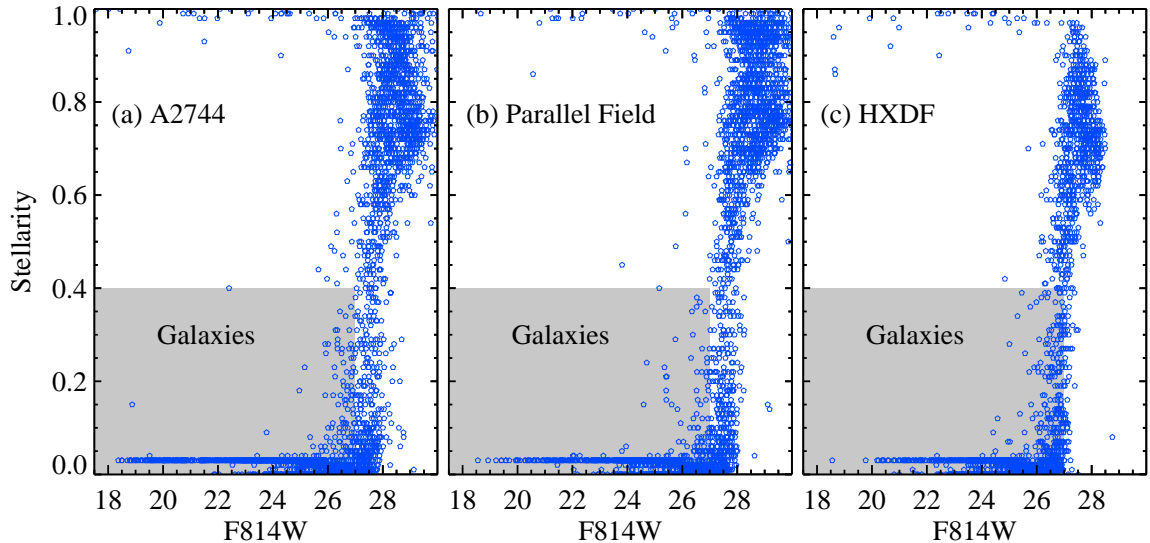}
\caption{Stellarity vs. F814W (Vega) magnitudes of the sources detected with SExtractor in the Abell 2744 field (a), the parallel field (b), and the HXDF (c). 
We selected the extended sources with stellarity $<0.4$ and F814W$<27.0$ to study the galaxy populations in Abell 2744. 
The sources with stellarity $>0.6$ are mostly point sources.
Note that the Abell 2744 photometry goes $\sim$1.0 mag deeper than the HXDF photometry.
}
\label{fig_stell}
\end{figure*}

\subsection{Detection of Point Sources}
{\bf Figure 1(b)} shows that diffuse light in and around bright galaxies occupies a significant fraction of the field. We removed diffuse light from the images before source detection. 
We made a smooth background image for each filter using ring median filters, and subtracted them from the original images in {\bf Figure 1(b)}. These background-subtracted images (a small section of which is shown in {\bf Figure 1(c)}) were used for the detection of point sources. 
Many point sources (marked with red circles) are clearly seen in the background-subtracted images ({\bf Figure 1(c)}), with most being 
GC and UCD candidates. A small number of extended sources are also seen in the same image, and these are galaxies.

\begin{figure*}
\centering
\includegraphics[scale=0.8]{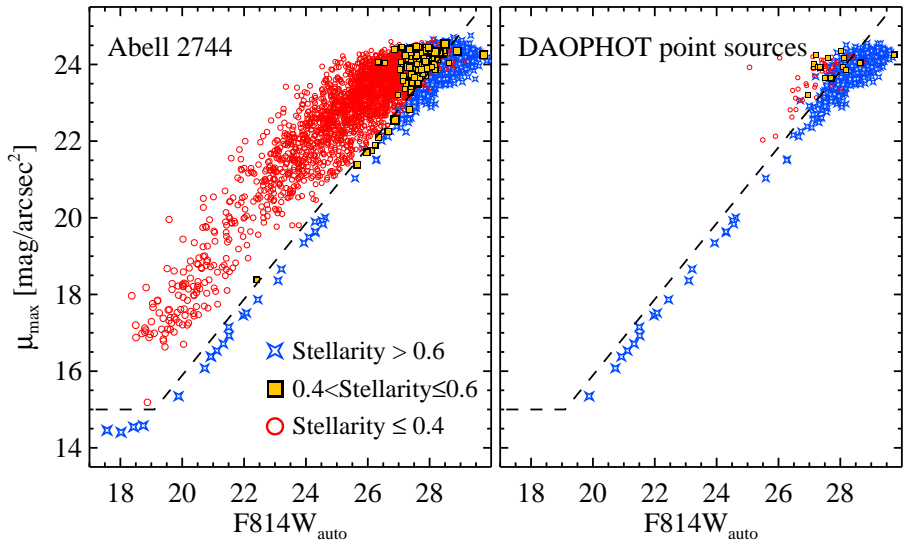}
\caption{The maximum surface brightness ($\mu(0)$) at the center of each source vs. F814W (AUTO magnitude in the Vega system) in Abell 2744, which is useful for separating extended sources from point sources. 
Circles, squares, and starlets represent the sources
with stellarity$<0.4$, $=0.4-0.6$, and $>0.6$, respectively.
(a) The sources detected with SExtractor.
(b) The point sources selected  with DAOPHOT in F814W images (before matching with F105W photometry).
The dashed line represents the boundary for distinguishing extended sources from point sources detected with SExtractor.
Note that the extended sources are well separated by the stellarity up to F814W$<27.0$ mag.
}
\label{fig_mumag}
\end{figure*}

We used  DAOPHOT \citep{ste87,ste94} to find point sources and derive the point spread function fitting photometry of all detected point sources.
We selected the point sources among the detected sources using the DAOPHOT sharpness parameters.
Then we calibrated 
the instrumental magnitudes of these sources to the Vega system 
 following the STScI webpage\footnote{http://www.stsci.edu/hst/acs/analysis/zeropoints},\footnote{http://www.stsci.edu/hst/wfc3/phot\_zp\_lbn}. 
 Note that \citet{ala13} adopted AB magnitudes for their photometry of the point sources in Abell 1689.  
Vega magnitudes for F814W are 0.424 mag brighter than AB magnitudes: F814W(Vega) = F814W(AB)$-0.424$. 

We performed artificial star experiments to estimate the completeness of our photometry for the point sources. We added a number of artificial star images to the HST images using IRAF/ADDSTAR, and carried out detection and photometry of the sources using the same procedures as done for the original images. 
We set the color of the artificial stars to be (F814W--F105W)$=0.6$. Then we calculated the ratio of the number of the detected stars and the number of the added star to estimate the recovery rate. {\bf Figure \ref{fig_completeness}} displays the results of the artificial star experiments. This figure shows
that the completeness of the point source photometry is 50\% at F814W$\approx 29.2$ mag for both the Abell 2744 field and the parallel field. It also shows that the mean differences in F814W magnitudes and (F814W--F105W) colors between input and output artificial stars are smaller than 0.05 mag for F814W$<29.2$ mag.

\subsection{Detection of Galaxies}

For the study of galaxies in Abell 2744, we used SExtractor \citep{ber96} to detect extended sources and derive their photometry in the  original  images. 
Stellarity parameters provided by  SExtractor are an excellent criterion to distinguish  extended sources from point sources in the images.
{\bf Figure \ref{fig_stell}(a)} displays  stellarity values versus F814W magnitudes
of all detected sources in the Abell 2744 field.
We selected the extended sources with stellarity smaller than 0.4 in the SExtractor photometry catalog, to produce a sample of galaxies.  For analysis of galaxies, we used only the sources brighter than
F814W$=27.0$ mag for which point sources and extended sources are clearly separated by stellarity values.
We adopted AUTO magnitudes of the sources provided by  SExtractor.
These extended sources are thought to be mostly galaxies.

For estimating the background contribution to the Abell 2744 field, we used two sets of data: data for the parallel field of Abell 2744 and data for the the Hubble Extreme Deep Field (HXDF) provided by \citet{ill13}.
We performed similar analysis of the images of the parallel field $5\farcm7$ west of the Abell 2744 field, as shown in {\bf Figure 1(a)}, in the same released data.
The data for the parallel field were obtained with similar exposure times, reaching the similar depth of photometry as the Abell 2744 field. 
Similarly, we analyzed also the F814W and F105W $2\farcm0 \times 2\farcm3$ images of the HXDF. 
Basic information of the HST data for the HXDF
is also included in {\bf Table \ref{tab_data}}.
The HXDF 
is located  far from the Abell 2744 field in the sky, while the parallel field is much closer to 
the Abell 2744 field. 
Thus, the HXDF images can be used to estimate the background contribution  to the Abell 2744 field. The parallel field may contain a small number of the Abell 2744 cluster members, which can be checked with the results of the HXDF.
The exposure times for the HXDF images are about a half of those for Abell 2744 images, but the source crowding and diffuse light level of large galaxies in the HXDF images are much less than those in the Abell 2744 images.

\begin{figure*}
\centering
\includegraphics[scale=0.75]{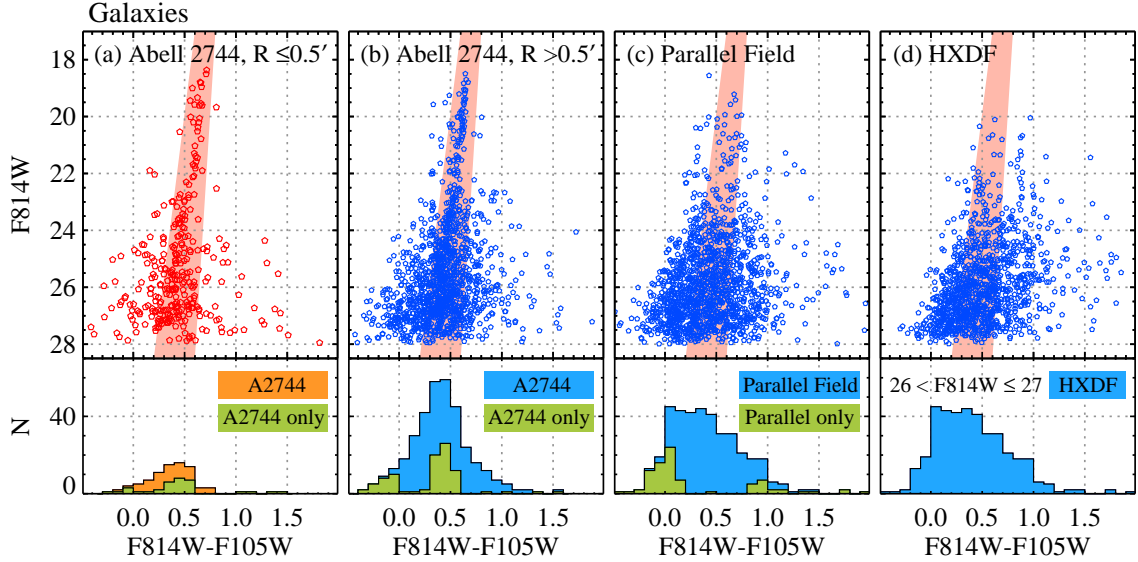} 
\caption{ (Upper panels) CMDs of the extended sources (with stellarity $<0.4$ and F814W$<28$ mag) in the Abell 2744 field, the parallel field, and the HXDF. Red and blue symbols represent the sources in the inner 
($R<0\farcm5$) and outer regions ($R>0\farcm5$), respectively, where $R$ is a projected angular distance from the center of the brightest galaxy, CN-2. 
Shaded regions denote the boundaries used for selecting the red galaxy sequence galaxies.
(Lower panels) Color distributions of faint galaxies with $26<F814W<27$ mag before (orange and cyan histograms) and after (green histograms) subtracting the background contribution with the HXDF.
 Note that the strong peak at (F814W--F105W)$\approx 0.45$  is due to the red sequence galaxies in Abell 2744.  The blue excess in the Abell 2744 field  is mainly due to the lensed images of background star-forming galaxies, while that in the parallel field is due to the star-forming galaxies in Abell 2744.
}
\label{fig_cmdext}
\end{figure*}

\begin{figure*}
\centering
\includegraphics[scale=0.8]{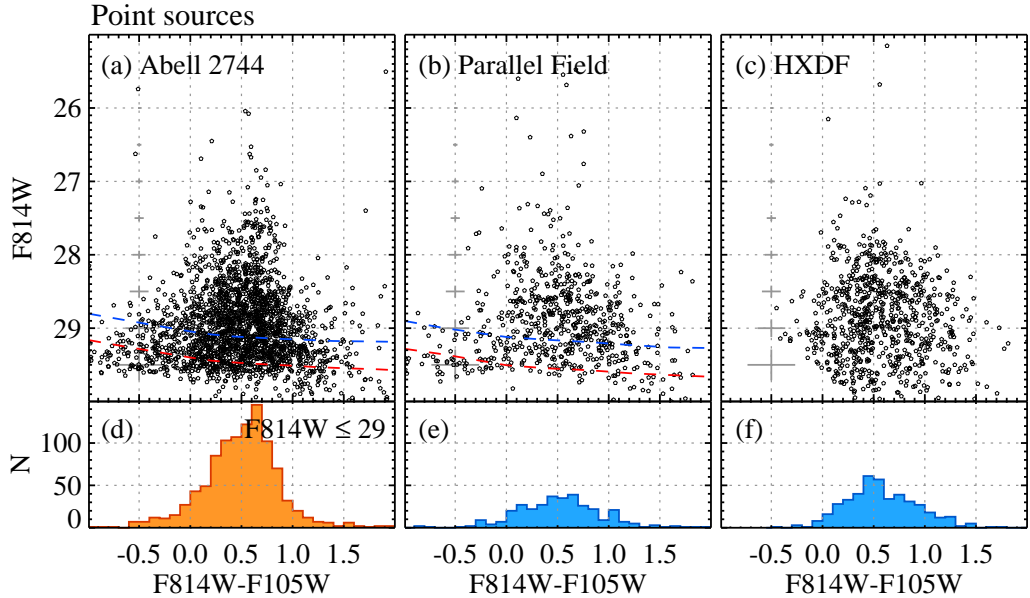} 
\caption{ (Upper panels) CMDs of the point sources detected with DAOPHOT, in the Abell 2744 field (a),
in the parallel field (b), and in the HXDF (c).
(Lower panels) Histograms of the bright point sources with F814W(Vega)$<29.0$ mag of each field. 
Blue and red dashed lines in the upper panels denote the completeness of 50\% and 20\%, respectively.
Note that the CMD of the point sources in Abell 2744 shows a dominant population in the vertical branch at (F814W--F105W)$\approx 0.5$, much stronger than those in the CMDs of the reference fields. The dominant population in Abell 2744 consists mostly of UCDs and bright GCs. 
}
\label{fig_cmdpoint}
\end{figure*}

The relation between surface brightness and integrated magnitudes has been sometimes used to distinguish between the point sources and the extended sources in the study of distant galaxies \citep{ban10, fer12, dep13}.
In {\bf Figure \ref{fig_mumag}(a)}, 
we plotted  the maximum surface brightness ($\mu(0)$, at the center of each source) versus F814W magnitudes of the sources derived using  SExtractor. We marked the extended sources selected using the condition of stellarity $<0.4$ with open red circles,
0.4 to 0.6 with yellow squares, and
$>0.6$ with blue starlet symbols.
This figure shows that point sources (with stellarity $>0.6$) follow a narrow sequence up to the faint magnitude at  F814W$\approx 27$ mag, while extended sources (with stellarity $<0.4$) follow another sequence with lower surface brightness and larger scatter. These two sequences overlap in the faint end at F814W$>27$ mag. Thus, the stellarity parameter is found to be very efficient at distinguishing between extended sources and point sources, as shown by the dashed line in the figure.
In the following analysis we use the stellarity condition ($<0.4$) to select the extended sources for the study of galaxies.
In {\bf Figure \ref{fig_mumag}(b)}, 
we plotted a  diagram similar to {\bf Figure \ref{fig_mumag}(a)},  but for the point sources selected using DAOPHOT. This figure shows that the position of the DAOPHOT point sources follows well the relation for the SExtractor point sources with stellarity $>0.6$. A small number of the faint DAOPHOT point sources with  27$<F814W<$28.0 mag have $0.4<$stellarity$<0.6$, and some with 26$<F814W<$28.0 mag have
stellarity$<0.4$. The stellarity values show a large spread at the faint level close to the detection limit so that sensitivity of the stellarity used for classification decreases significantly.
Thus we use the DAOPHOT point sources for the analysis 
in the following.

\section{RESULTS}

\begin{figure}
\centering
\includegraphics[scale=0.65]{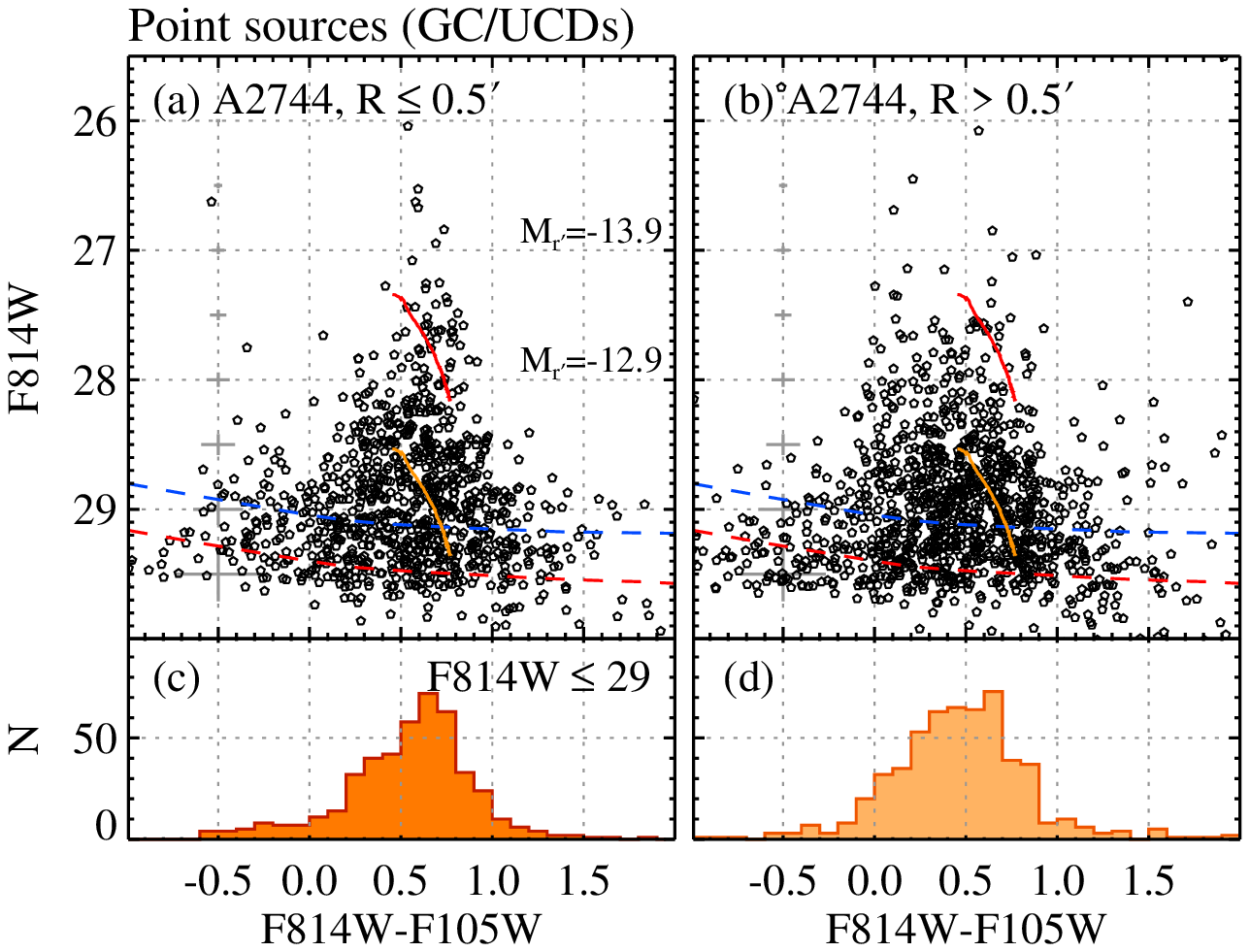} 
\caption{ CMDs of the point sources detected using DAOPHOT, in the inner region ($R\leq 0\farcm5$) (a) and in the outer region ($R>0\farcm5$) (b) of Abell 2744. 
Curved yellow and red lines represent SSP models for age of 10 Gyr and masses of $M=10^7 M_\odot$ and $M=3\times 10^7 M_\odot$, respectively.
 SSP models cover  a metallicity range 
 from [Fe/H] = --2.4 to 0.0 (left to right). 
Blue and red dashed lines in the upper panels denote the completeness of 50\% and 20\%, respectively.
The histograms in (c) and (d) denote the color distribution of the bright point sources with $F814W<29.0$ mag.
Note that the peak color of the sources is consistent with the model for low metallicity [Fe/H]$\approx -2$.
Errorbars at (F814W--F105W)$=-0.5$ represent the mean errors for given magnitudes.
}
\label{fig_cmdpointinout}
\end{figure}


\begin{figure*}
\centering
\includegraphics[scale=0.8]{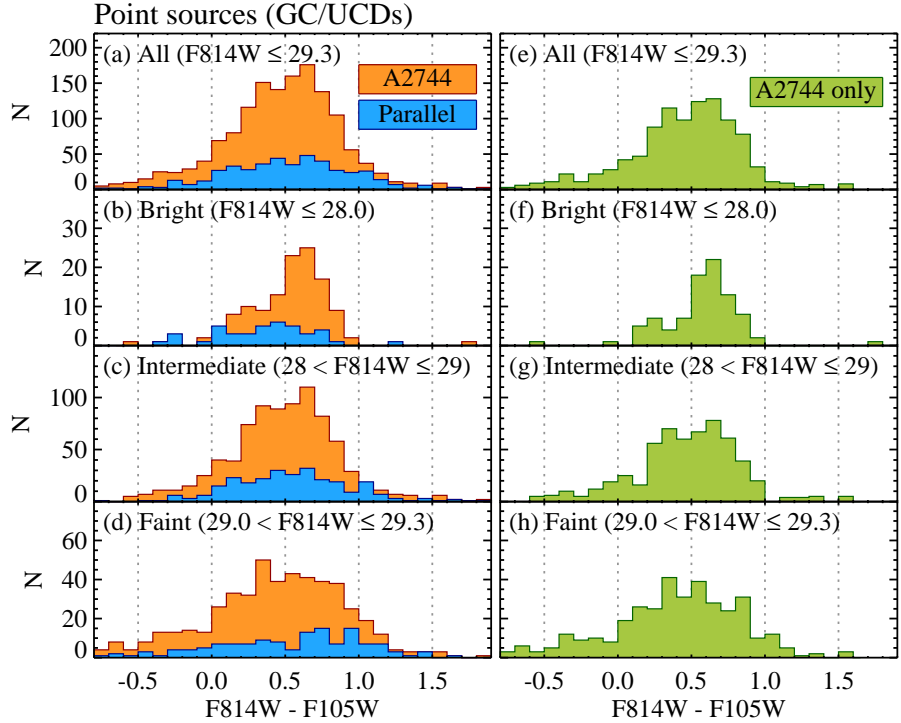} 
\caption{Color histograms of all point sources in Abell 2744 and the parallel field before (left panels)  and after (right panels) background subtraction using the data for the parallel field.
Top, middle 1 and middle 2, and bottom panels are for all sources ($26.0<$F814W$<29.3$ mag), 
bright sources  ($26.0<$F814W$<28.0$ mag),
intermediate luminosity sources   ($28.0<$F814W$<29.0$ mag),
and faint sources   ($29.0<$F814W$<29.3$ mag),
  respectively.
}
\label{fig_cdfpoint}
\end{figure*}


\begin{figure*}
\centering
\includegraphics[scale=0.8]{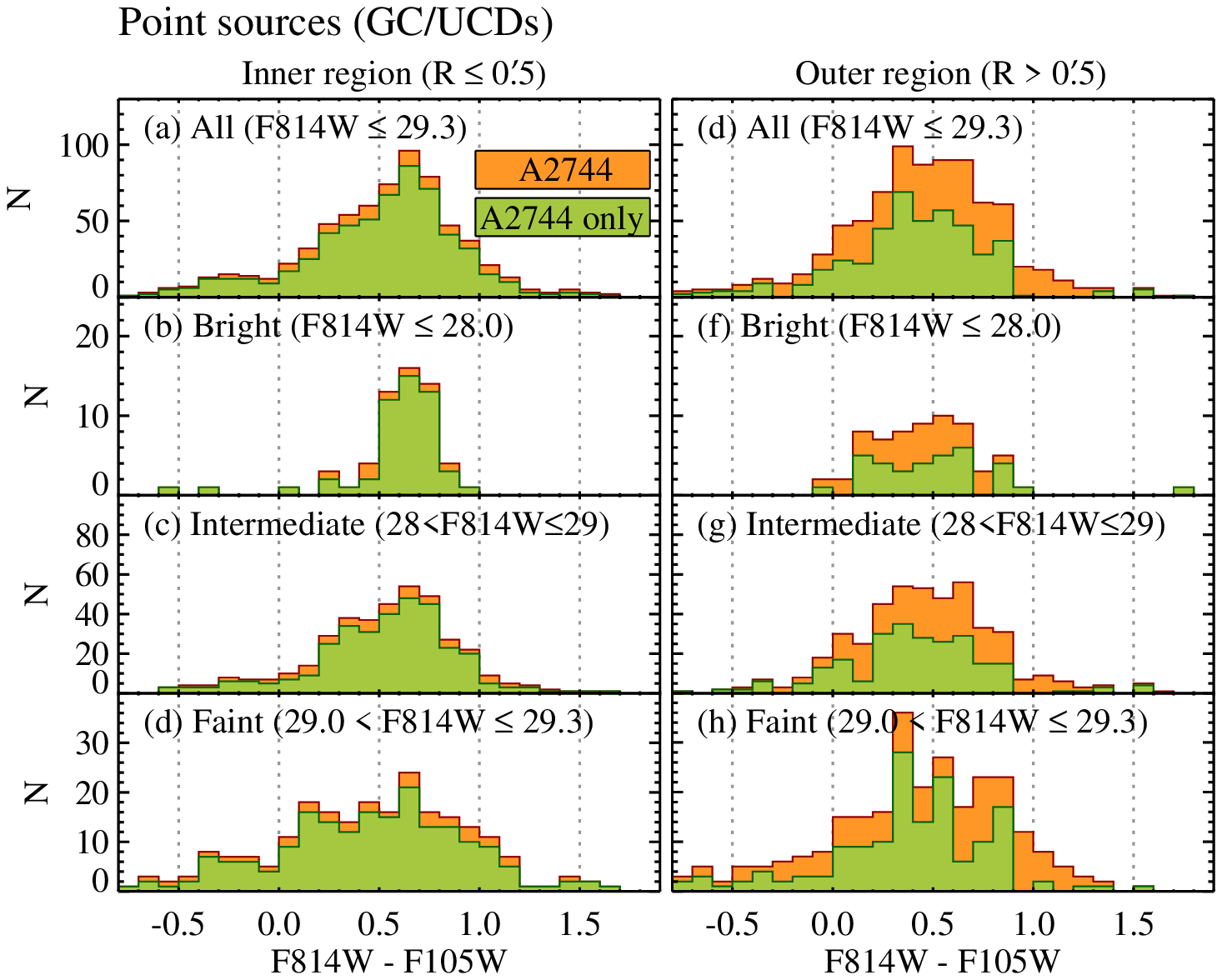} 
\caption{
Same as Fig. \ref{fig_cdfpoint}(right panels) except for 
the point sources in the inner region ($R\leq 0\arcmin.5$) (left panels) and  in the outer region ($R>0\arcmin.5$) (right panels) of Abell 2744.
}
\label{fig_cdfpoint2}
\end{figure*}

\subsection{CMDs of the Galaxies}

The point sources detected in the HST images are mostly unresolved GCs and UCDs, and the extended sources are mainly galaxies.
We present  the CMDs of the extended sources to study galaxies in this section, and
those of the point sources to investigate the GCs and UCDs in Abell 2744 in the following section. 
{\bf Figure \ref{fig_cmdext}} displays F814W--(F814W--F105W) CMDs of the extended sources (having stellarity $<0.4$) brighter than F814W$=28.0$ mag ($M_{r'} = -12.9$ mag) detected in the images of the Abell 2744 field, the parallel field, and the HXDF. 
In {\bf Figure \ref{fig_cmdext}(a,b)} we plotted, with red and blue symbols, respectively, 
the sources in the inner region at $R<0\farcm5$ (= 131 kpc at $z=0.308$) 
and those in the outer region at $R>0\farcm5$,  where $R$ is the projected distance from the center of the brightest galaxy, CN-2.

The most remarkable feature in the CMD of the extended sources in Abell 2744 is the existence of a narrow red sequence, the brightest of which extends to F814W $\approx 18.5$ mag and (F814W--F105W)$\approx 0.7$ ($M_{r'} = -22.4$ mag, $M_I = -23.1$ mag). The bright part of this sequence is consistent with the old findings based on ground-based data by \citet{cou87}, but the sequence in this study goes much fainter. 
This red sequence is 
more distinguishable in the inner region than in the outer region, indicating that most of the galaxies in this sequence are the members of Abell 2744.  
This sequence becomes broader as the sources become fainter, due to increasing photometric errors and the increasing fraction of blue galaxies.
This red sequence is fit well by a linear relation,
F814W $= (-32.7\pm0.3)($F814W--F105W$)+(39.8\pm0.2)$ for
the bright range ($18.5<$F814W$<26.0$ mag). The faint part of this red sequence  becomes
almost vertical at (F814W--F105W)$\approx 0.4$.
There are a number of sources in the blue side of the red sequence of Abell 2744, showing the Butcher-Oemler effect.
The red sequence is not obvious in the CMDs of the parallel field and the HXDF, showing again that the red sequence galaxies in the  Abell 2744 field are mostly the members of the cluster.

To investigate the properties of faint galaxies in Abell 2744, we derived the color distributions of the faint extended sources with 
$26.0<$F814W$<27.0$ mag ($-14.9<M_{r'} < -13.9$ mag)  in each field, 
plotting them with the orange and cyan histograms in 
{\bf Figure \ref{fig_cmdext}} (lower panels).
We subtracted the background contribution using the data from the HXDF, plotting the results with the green histograms. 
Two interesting features can be found in this figure.
First, the background-subtracted color distribution of the Abell 2744 field shows a narrow red component with $0.3<$(F814W--F105W)$<0.6$ whose peak is at (F814W--F105W)$=0.41\pm0.01$. This is not seen in the case of the parallel field.
This shows that most of these red faint galaxies must be the members of Abell 2744,
and that the parallel field contains few red sequence galaxies.
The colors of these faint galaxies in Abell 2744  are similar to those of red dwarf galaxies found in the local universe.
Second, both the Abell 2744 field and the parallel field show an excess of faint blue galaxies at  $-0.3<$(F814W--F105W)$<0.1$. 
The number of these sources in the Abell 2744 field is $N=32$, which is about a half of that in the parallel field, $N=54$. 
Thus the number density of these galaxies is higher in the parallel field than in the Abell 2744 field.
We checked the images of these blue faint sources, finding that most of them are 
star-forming dwarf galaxies. 
Thus these blue faint galaxies are probably star-forming galaxies that are located in the outskirts of Abell 2744, falling toward the cluster center.

We marked the boundary of the red sequence following the position of the red galaxies in the inner region of Abell 2744 in the CMD by the shaded region.
The sources located within the red sequence boundary 
are most likely the members of Abell 2744. We selected these red sequence galaxies  to investigate the properties of the cluster member galaxies in the following section.

\subsection{CMDs of the GCs and UCDs}
The CMDs of the point sources in each field are displayed in {\bf Figure \ref{fig_cmdpoint}}, which includes also the color histograms of the sources brighter than F814W $=29.0$ mag.
 We also plotted the CMDs of the point sources in the inner ($R \leq 0\farcm5$) and outer ($R>0\farcm5$) regions of the Abell 2744 field in {\bf Figure \ref{fig_cmdpointinout}}.
A remarkable feature in the CMDs of Abell 2744 is the presence of a rich population of point sources in the broad vertical feature at $0.0<$(F814W--F105W)$<1.0$ and $26.0<$F814W$<29.5$ mag ($-14.9 < M_{r'}  < -11.4$ mag). 
These sources are mostly fainter than F814W = 27.5 ($M_{r'}  \approx -13.4$) mag, and their color distribution shows a strong peak at (F814W--F105W) $\approx 0.6$. 

In {\bf Figure \ref{fig_cmdpointinout}}, we also plotted simple stellar population (SSP) models for an age of 
10 Gyr, masses of $M=10^7~ M_\odot$ and
$M=3\times 10^7~ M_\odot$, 
 and metallicity  [Fe/H] = --2.4 to 0.0 (from blue to red),
based on the Padova models \citep{gir00}. 
The peak color in the color distribution of the 
 point sources in {\bf Figure \ref{fig_cmdpointinout}} 
is consistent with the SSP models for low metallicity (Fe/H]$<-2.0$). The magnitudes and colors of these point sources (F814W $\approx 28.5$ mag and (F814W--F105W) $\approx 0.5$) 
are, on average, consistent with $M\approx 10^7 M_\odot$ models for low metallicity.
This indicates that these point sources are mostly UCDs and massive GCs  that have much higher masses than the massive GCs in the Milky Way Galaxy, and that they are metal-poor.

We divided all point sources with $26.0<$F814W$<29.3$ mag
into three groups according to their magnitudes:
a bright group ($26.0<$F814W$<28.0$ mag, $-14.9 < M_{r'}  \leq  -12.9$ mag),
an intermediate-luminosity group ($28.0<$F814W$<29.0$ mag, $-12.9 < M_{r'}  \leq -11.9$ mag),
and a faint group ($29.0<$F814W$<29.3$ mag, $-11.9 < M_{r'}  \leq -11.6$ mag). 
The color distributions for these groups  in each field 
are plotted in  {\bf Figure \ref{fig_cdfpoint}}.  
The color distributions for Abell 2744 (orange histograms) in the left panels of the figure follow a Gaussian distribution, while those for the parallel field (blue histograms) are relatively flat. This shows that the point sources in the parallel field are mostly background sources.
Thus, the contribution of the point source members of Abell 2744 is minor in the parallel field. The HXDF data is not as deep as the data for Abell 2744 for the study of the point sources. 
Therefore, we subtracted the contribution due to the background using the color distribution for the parallel field, and plotted the results with green histograms in {\bf Figure \ref{fig_cdfpoint}(right panels)}.
The background-subtracted color distribution of all point sources  shows a strong concentration with a peak at (F814W--F105W)$\approx 0.5$.
The color distribution of the bright group shows a strong concentration  at $0.5<$(F814W--F105W)$<0.8$ with a peak at (F814W--F105W)$\approx 0.6$, 
while that of the intermediate-luminosity group shows a peak at a slightly bluer color, (F814W--F105W)$\approx 0.5$. 

We display the color distributions of the point sources in the inner ($R\leq 0\arcmin.5$) and outer ($R>0\arcmin.5$) regions of Abell 2744  before and after background subtraction with the parallel field data in  {\bf Figure \ref{fig_cdfpoint2}}. 
The peak colors of the inner region are slightly redder than those of the outer region. 
The peak color for the inner region is (F814W--F105W)$= 0.65$ mag, 
which is $\sim$0.1 mag 
redder than that for the outer region.

These results imply that the bright sources (UCDs) are, on average, redder (more metal-rich) than the intermediate-luminosity group sources (bright GCs), and that the fraction of redder GCs/UCDs is higher in the inner region than that in the outer region. 
 This trend has been seen also in the CMDs of the local galaxy clusters like the Coma cluster (see \citet{chi11} and their Figure 8).
The color distribution of the faint group is broader than those of the brighter groups,
which is mainly due to increasing photometric errors.

The faint group and the intermediate-luminosity group show also a small population of blue sources with $-0.4<$(F814W--F105W)$<0.0$. 
We checked the morphology of these sources in the images. These faint blue point sources are mostly compact star-forming dwarf galaxies like blue compact dwarfs (BCDs), and
a few of them are considered to be lensed images of background blue galaxies.

\begin{figure}
\centering
\includegraphics[scale=0.85]{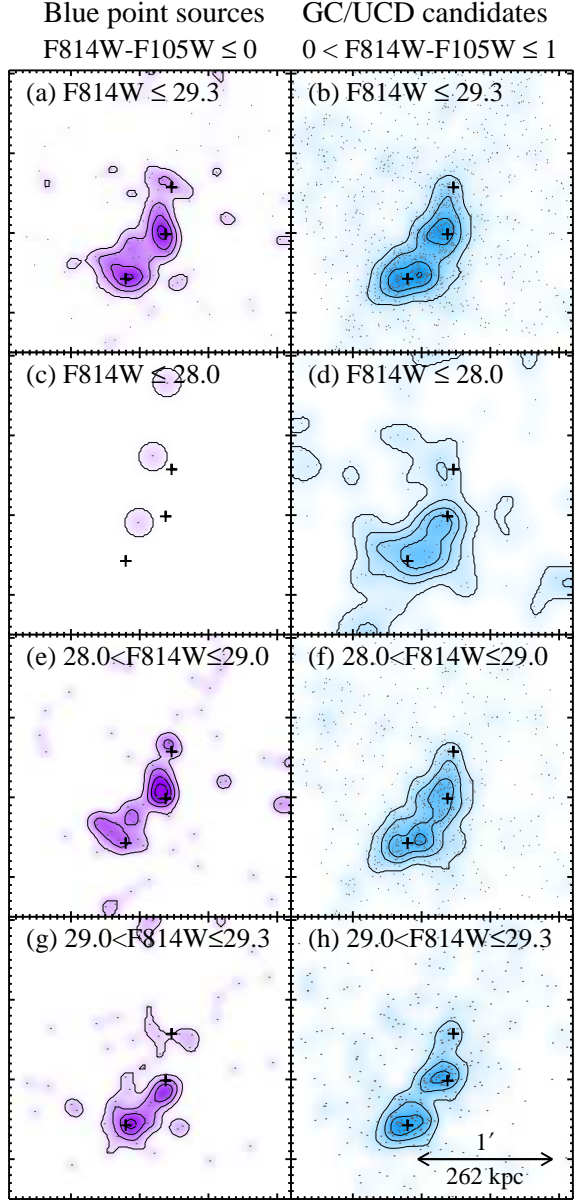} 
\caption{ 
Spatial distributions of 
blue point sources ($-0.5<$(F814W--F105W)$<0.0$) and  all GC/UCD candidates ($0.0<$(F814W--F105W)$<1.0$) in Abell 2744.
Top, middle 1 and middle 2, and bottom panels are for all sources ($26.0<$F814W$<29.3$ mag), 
bright sources  ($26.0<$F814W$<28.0$ mag),
intermediate luminosity sources   ($28.0<$F814W$<29.0$ mag),
and faint sources   ($29.0<$F814W$<29.3$ mag),
  respectively.
}
\label{fig_spatpoint}
\end{figure}

\subsection{Spatial Distributions of the GCs and UCDs}

%
%
We selected, for the GC and UCD candidates, the point sources with $26.0<$F814W$<29.3$ ($-14.9<M_{r'}  < -11.6$)  mag and $0.0<$(F814W--F105W)$<1.0$ in the DAOPHOT photometry catalog of Abell 2744 for further analysis.
We divided the GC/UCD sources 
into three subgroups 
according to their magnitudes as in the previous section.
We also  selected a blue group of the point sources  that have $-0.5<$(F814W--F105W)$\leq 0.0$.

{\bf Figure \ref{fig_spatpoint}} 
shows the spatial distribution of the selected sources in Abell 2744,
and their number density maps in contours
(the blue sources in the left panels and the GC/UCDs in the right panels).
This figure shows two notable features. First, 
it is seen clearly that the spatial distribution of the GCs and UCDs shows a strong central concentration around bright galaxies 
close to the center of the southern core in Abell 2744. The peak with the highest number density  is close to the center of the brightest galaxy, CN-2. 
The central concentration is seen in both the bright and faint groups. It is noted that the extent of their spatial distribution reaches out to $0\farcm5$ ($\approx 131$ kpc) and further, indicating that they include not only galaxy GCs but also intracluster GCs.
Second, 
the spatial distribution of the blue faint sources 
also shows a central concentration around bright galaxies, although their number density is much lower than that of the GC/UCDs.
These blue sources are mostly star-forming galaxies,
and some of them are gravitationally-lensed sources of background star-forming galaxies.

\begin{figure}
\centering
\includegraphics[scale=0.9]{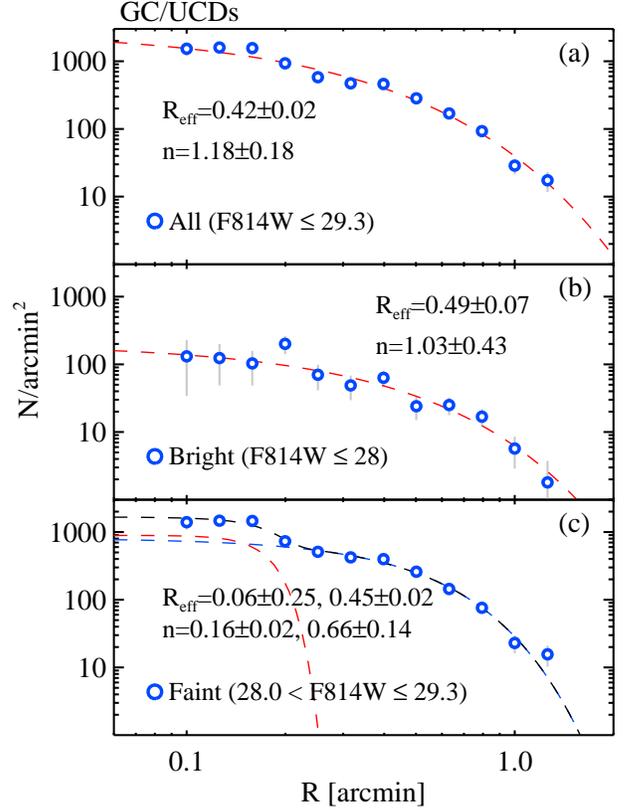} 
\caption{
Radial number density profiles of the GC/UCD candidates in Abell 2744. We subtracted background contributions estimated using the mean value of the outermost regions at $R=1\farcm4$ in the Abell 2744 field:
(a) all sources with F814W$\leq 29.3$ mag; (b) the bright sources ($26<$F814W$\leq 28.5$ mag); and (c) the faint sources ($28.5<$F814W$\leq 29.3$ mag).
Dashed lines represent the fitting results with a S{\'e}rsic law with an index $n$ and effective radius $R_{\rm eff}$.
}
\label{fig_raddenpoint}
\end{figure}

\begin{deluxetable}{lcccccccc}
\tablecolumns{9}
\tablewidth{0pc}
\tablecaption{A Summary of S{\'e}rsic Law Profile$^a$ Fits on GC/UCDs in Abell 2744}
\tablehead{
\colhead{ } & \colhead{$I_{\rm eff}$} & \colhead{$R_{\rm eff}$} & \colhead{$n$}}

\startdata
All GC/UCDs		& $367.4\pm35.3$ & $0\farcm41\pm0\farcm02$	& $1.02\pm0.15$   \\
Bright GC/UCDs		& $38.0\pm15.8$ & $0\farcm39\pm0\farcm08$	& $1.55\pm0.82$   \\
Faint GC/UCDs (outer)		& $294.5\pm27.4$ & $0\farcm45\pm0\farcm02$	& $0.58\pm0.11$   \\
Faint GC/UCDs (inner)		& $771.2\pm284.8$ & $0\farcm06\pm0\farcm23$	& $0.16\pm0.02$   \\
\enddata
\label{tab_sersic}
\tablenotetext{a}{Using Equation (1) of \citet{gra05}, $I(R)=I_e$ exp$\{-b_n[(R/R_e)^{1/n}-1]\}$, where $b_n=1.9992n-0.3271$.}
\end{deluxetable}

\subsection{Radial Number Density Profiles of the GCs and UCDs}


{\bf Figure \ref{fig_raddenpoint}} displays the radial number density profiles of the GCs and UCDs in Abell 2744.
We adopted the center of CN-2, which is close to the position of the highest number density, 
as a reference center for deriving the radial number density profiles.
We subtracted the contribution due to background sources using the number density of the point sources in the outermost region at $R>1\arcmin.4$ (367 kpc). 
The radial number density profile of the GCs and UCDs in {\bf Figure \ref{fig_raddenpoint}(a)} shows a slight drop in the central region at $R\approx 0\farcm2$ (52 kpc), decreases slowly as the clustercentric distance increases until at $R\approx 0\farcm7$ (183 kpc), and decreases more rapidly thereafter until the outer boundary of Abell 2744,  at $R\approx 1\farcm3$ (341 kpc).  
A break at  $R \approx 0\farcm2$ indicates 
that the radial profiles consist of two components:
a narrow one in the central region and a broad one in the outer region. 

We divided the sample of GCs and UCDs into two subgroups according to their magnitudes:
a bright group with F814W$\leq 28.0$ mag  (UCDs), and
a faint group with $28.0<$F814$\leq 29.3$ mag  (GCs).
{\bf Figure \ref{fig_raddenpoint}(b,c)} show little difference in the shapes of the radial density profile between the bright group and the faint group except for the central region where the faint group shows a slight excess.
These radial profiles do not follow a power law. Instead, they are described well by a S{\'e}rsic law \citep{ser63, gra05}, as in the case of NGC 6166, a cD galaxy in Abell 2199 \citep{har16}.
Fitting the data for the region at $0\farcm1<R<1\farcm3$ with a S{\'e}rsic law with an index $n$ and effective radius $R_{\rm eff}$, we obtain the results as follows and as listed in {\bf Table \ref{tab_sersic}}:
$n=1.18\pm0.18$ and  $R_{\rm eff} = 0\farcm42\pm0\farcm02$ for the entire sample;
$n=1.03\pm0.43$ and $R_{\rm eff} = 0\farcm49\pm0\farcm07$ for the bright group.
The data for the faint group is better fit with a double S{\'e}rsic law: 
$n=0.66\pm0.14$ and $R_{\rm eff} = 0\farcm45\pm0\farcm02$ 
for the outer component, and 
$n=0.16\pm0.02$ and $R_{\rm eff} = 0\farcm06\pm0\farcm25$
for the central component.
The central component represents mostly galaxy GCs belonging to CN-2, while the latter contains intracluster GCs in Abell 2744.
Thus, the values of the S{\'e}rsic index derived for the entire radial range are similar to those of the exponential disk law, and their differences between the bright and faint groups are not significant except for in the central region.

\begin{figure}
\centering
\includegraphics[scale=0.9]{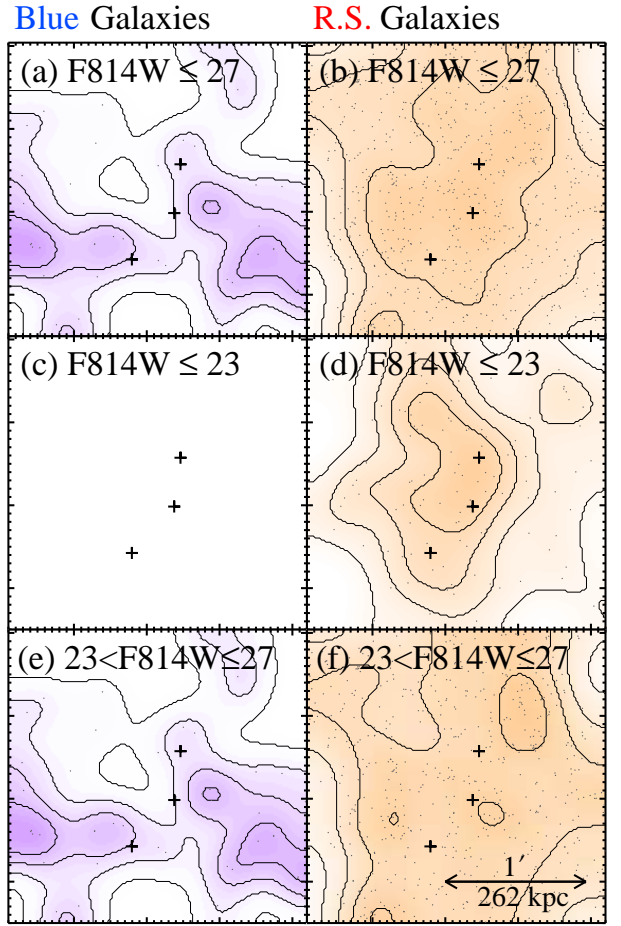} 
\caption{
Spatial distributions of 
blue galaxies ($-0.5<$(F814W--F105W)$\leq 0.0$) (left panels) and red sequence galaxies (right panels) with $18.0<$F814W$\leq 27.0$ mag (top panels).
Middle and bottom panels represent, respectively, bright sources ($18.0<$F814W$\leq 23.0$ mag), 
and faint sources ($23.0<$F814W$\leq 27.0$ mag).   
}
\label{fig_spatext}
\end{figure}

\begin{figure}
\centering
\includegraphics[scale=1.0]{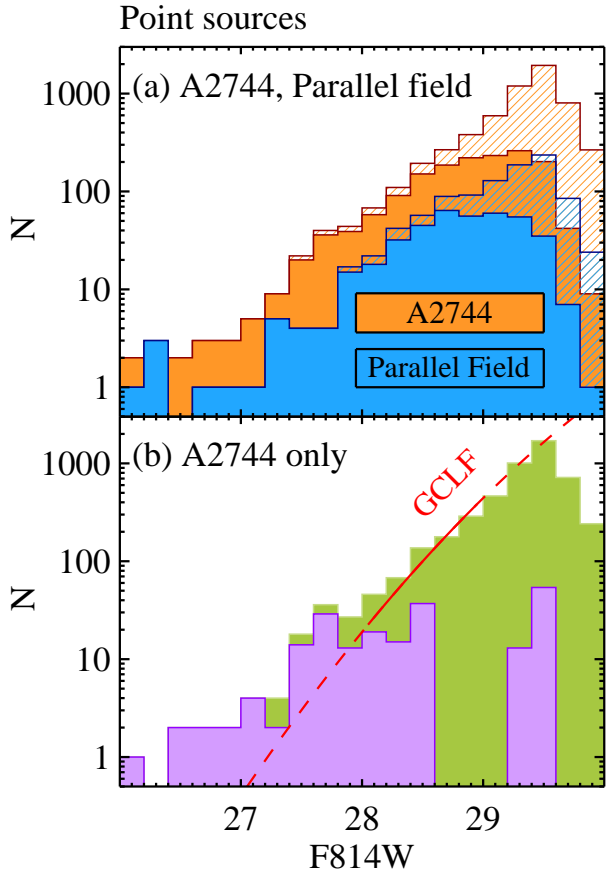} 
\caption{
LFs of the point sources with $0.0<(F814W-F105W)<1.0$  in the  Abell 2744 field (orange color in (a)) and in the parallel field (blue color in (a)). 
Filled and hatched histograms denote the LFs derived after completeness correction.
The green histogram in (b) denotes the completeness-corrected LF after the background contribution was subtracted using the data for the parallel field. Thus this histogram represents the LF of the GC/UCDs in Abell 2744.
The red curved line represents, as a reference for the GCLF, a Gaussian function with $\sigma=1.2$ and F814W$({\rm max}) = 33.0$ ($M_{r'} ({\rm max}) \approx -7.9$) mag,
shifted to match the data for F814W$=28.5$ mag for (b). 
The violet histogram in (b) denotes the LF after the contribution of GCs was subtracted from the green histogram, showing the LF of the excess UCDs.
}
\label{fig_lfpoint}
\end{figure}

\subsection{Spatial Distributions of the Galaxies}
We selected two groups based on the location of the extended sources in the CMD.
We chose, as the cluster member red sequence galaxy candidates, the extended sources located inside the red sequence boundary as marked in the CMD. Then we  selected a  blue group including the extended sources with  blue colors, $-0.5<$(F814W--F105W)$<0.0$. 
These blue galaxies are mostly fainter than F814W$=24.0$ ($M_{r'} = -16.9$) mag.  
This blue group may contain mostly
star-forming dwarf galaxies in Abell 2744,
 while those in the red sequence group are mostly early-type members.
 
{\bf Figure \ref{fig_spatext}} 
shows the spatial distribution of the selected sources,
and their number density maps in contours.
The spatial distribution of the red sequence galaxies shows a weak central concentration toward the center of the Abell 2744 field. 
The degree of central concentration is stronger for the bright galaxies with F814W $<23$ mag.
The number density of the faint galaxies varies much more slowly than that of the bright group.  
It is noted that the central concentration of the red faint galaxies is much weaker than that of the GCs and UCDs seen in {\bf Figure \ref{fig_spatpoint}}.
On the other hand, the spatial distribution of the blue galaxies show little central concentration, indicating no relation with the brightest galaxies in Abell 2744. However their number density is too low to have any
statistical significance.

\subsection{LFs of the GCs and UCDs}

We selected  red point sources with $0.0<$(F814W--F105W)$<1.0$ (corresponding to the color range of the GCs)  in Abell 2744 and the parallel field. Then we derived F814W-band LFs of these sources, as displayed in  {\bf Figure \ref{fig_lfpoint}(a)}.  
The filled histograms and hatched histograms in the figure denote the LFs before and after completeness correction, respectively.  
The completeness-corrected LF of these point sources 
increases as the magnitude gets fainter until F814W$\approx 29.5$ mag, and decreases significantly at F814W$\geq 29.5$ mag. The completeness is only 20\% at F814W$\approx 29.5$ mag. Thus the turnover at
F814W$\geq 29.5$ mag is probably due to incompleteness in our completeness correction for this faint level.
The LF of the red point sources  in the parallel field shows a similar trend, but in a lower fraction compared with that of the Abell 2744 field. 

We subtracted the contribution due to background sources using the LFs of the point sources in the parallel field, 
and plotted the net completeness-corrected LF by the green histogram in {\bf Figure \ref{fig_lfpoint}(b)}. 
This net LF can be considered to be the LF of the GCs and UCDs in Abell 2744. 
The net LF of the GCs and UCDs
increases rapidly from F814W$\approx 27.0$ mag ($M_{r'}  \approx -13.9$ mag) as the sources get fainter, 
becoming almost flat at  F814W$\approx 27.5$ mag ($M_{r'}  \approx -13.4$ mag). 
Then it increases again, 
reaching a maximum at F814W$\sim 29.5$ mag ($M_{r'}  \approx -11.4$ mag).
Then it shows a turnover, which is due to the incompleteness of our completeness correction. 
Thus, the net LF of the GCs and UCDs 
shows a break at F814W$\approx 28.0$ mag ($M_{r'}  \approx -12.9$ mag).
This break indicates that the net LF of the GCs and UCDs consists of
two distinguishable components: a flat one at the bright side,  
F814W$\lesssim28.0$ mag, 
and a steep one at the faint end, F814W$\geq 28.0$ mag.
The bright component is dominated by the UCDs, while the faint component may show the bright part of the GC LF.

As a reference for the LF of the GCs in the Milky Way Galaxy \citep{rej12,lee16}, we overlaid, by a red line in {\bf Figure \ref{fig_lfpoint}(b)}, a Gaussian function with $\sigma=1.2$ and  F814W$({\rm max}) = 33.0$ ($M_{r'} ({\rm max}) = -7.9$), arbitrarily shifted
vertically to match the data for $28.0<$F814W$<29.0$ mag  
in the net LF where photometric incompleteness is not significant.
It is noted that the LF for $28<$F814W$<29.5$ mag is well matched by the Gaussian function.
Then we subtracted the contribution of the normal GCs from the net LF using this Gaussian function, and plotted the resulting LF by the violet histogram in {\bf Figure \ref{fig_lfpoint}(b)}. 
This LF represents the LF of the UCDs only.
While a small number of the UCDs ($N\approx 10$) are brighter than F814W = 27.4 ($M_{r'} = -13.5$) mag, most of them have magnitudes F814W $= 27.4 \sim 28.6$ ($M_{r'} = -13.5 \sim -12.3$) mag. 
The number of sources with F814W$<28.6$ mag in the UCD LF is $N=147\pm26$. 
From the integration of the GCLF, we obtained a value of $N_{GC} = 385,044\pm24,016$, 
which is a rough estimate of the total number of the GCs in the Abell 2744 field. This value is 
even larger than that for Abell 1689, $N_{GC}=162,850^{+75,450}_{-51,310}$, given by \citet{ala13}. Thus Abell 2744 hosts the largest number of GCs among the known systems.
 Note that this conclusion is based on the assumption the LF of the GCs in Abell 2744 is the same as in the Milky Way.

\begin{figure}
\centering
\includegraphics[scale=0.9]{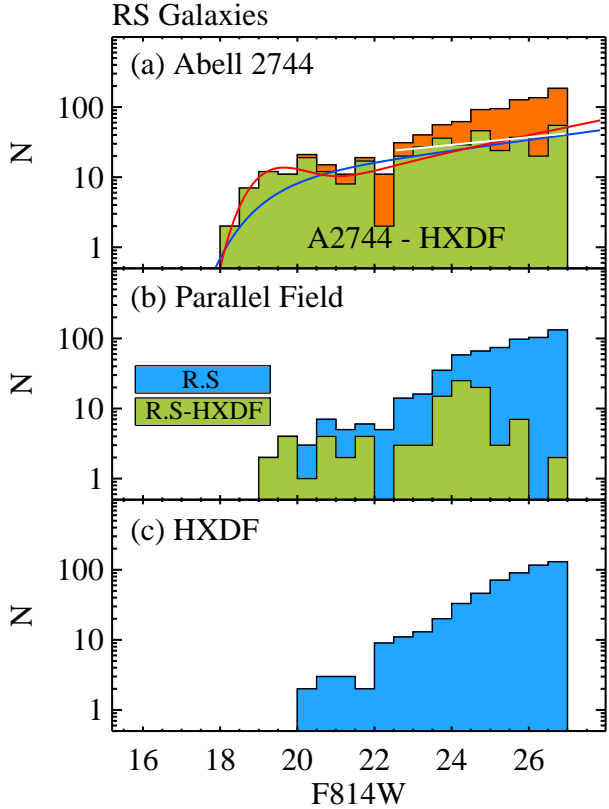} 
\caption{
LFs of  
the red sequence galaxies (orange and cyan histograms)  
in the Abell 2744 field (a),
in the parallel field (b), and in the HXDF (c).
Green histograms in (a) and (b) denote the LFs of the red sequence galaxies after the background contribution was subtracted using the HXDF data. 
Note that the LF of Abell 2744 is composed of two main components: a bright one and a faint one.
The blue and red lines represent the fits with a single Schechter function (with a power law index, $\alpha =-1.19\pm0.04$),
 and 
a combination of a Schechter function and a power law (with a power law index for the faint component, $\alpha =-1.21\pm0.76$), 
respectively.
The white line denotes a power law fit (with an index,
$\alpha =-1.14\pm0.08$), for the faint end 
($22.5<$F814W$\leq27.0$ mag) only.
}
\label{fig_lfext}
\end{figure}

\begin{deluxetable*}{lcccccccc}
\tablecolumns{9}
\tablewidth{0pc}

\tablecaption{A Summary of Fitting the Galaxy LF for Abell 2744}
\tablehead{
\colhead{ } & \colhead{$\Phi_b^*$} & \colhead{$M_b^*$} & \colhead{$\beta$} & \colhead{$\Phi_f^*$} & \colhead{$M_f^*$} & \colhead{$\alpha$}}
\startdata
Schechter function		& $11.0\pm2.0$ & $19.0\pm0.0^a$	& & & & $-1.19\pm0.04$   \\
Schechter function plus Power law$^b$		& $8.8\pm3.7$ & $20.2\pm0.4$	& $-1.28\pm0.08$ & & $20.7\pm0.3$ $(M_t)$ & $-1.21\pm0.76$  \\
Power law & & & & & &$-1.14\pm0.08^c$
\enddata
\label{tab_schechter}
\tablenotetext{a}{A fixed value.}
\tablenotetext{b}{As in \citet{ban10}.}
\tablenotetext{c}{Fitting only for the faint component with $22.5 < $F814W$ \leq 27.0$
($-18.4<M_{r'} \leq -13.9$) mag.}
\end{deluxetable*}

\subsection{LFs of the Red Sequence Galaxies}

{\bf Figure \ref{fig_lfext}} displays the F814W-band LFs
of the galaxies inside the red sequence boundaries in Abell 2744 (orange histogram), the parallel field (cyan histogram), and the HXDF (cyan histogram). 
We subtracted the contribution due to background sources using the LFs of the extended sources in the HXDF that were selected using the same criteria as for Abell 2744, plotting the resulting LFs by the green 
histograms.
The net LF of the red sequence galaxies in the parallel field shows only a small population of the member galaxies, which is negligible compared with that of the Abell 2744 field.

The net Lf of the red sequence galaxies in Abell 2744 shows a dip at F814W $\approx 22.3$ ($M_{r'} \approx -18.6$) mag, indicating a presence of two major components: a bright one and a faint one.
The bright component appears to be approximately Gaussian with a peak at F814W$\approx 20.0$ ($M_{r'}  \approx -20.9$) mag. The faint component follows a relatively flat power law, covering  $22.5 \lesssim $F814W$\lesssim  27$ ($-18.4 \lesssim M_{r'}  \lesssim  -13.9$) mag. 
The faint component represents a population of dwarf galaxies.
%
 
For the analysis of the galaxy LFs, either a single Schechter function or a combination of two functions has been used traditionally \citep{sch76,lov97,hil03,pop06,bar07,ban10,mor15}. 
The Schechter function, 
$\Phi (L) dL = \Phi^* (L/L^*)^\alpha$ exp$(-L/L^*) dL$,   is given in terms of magnitude $M$ by

$\Phi (M) dM = (0.4 \ln 10) \Phi^* 10^{-0.4(M-M^* )(\alpha+1)}$ exp$(-10^{-0.4(M-M^* )}) dM$ 

\noindent where
 $\Phi^*$, $M^*$, and $\alpha$ are fitting parameters to be derived from data \citep{sch76}.
 %

 
 \noindent We fit the galaxy LF with a Schechter function with a fixed value for F814W$^* =19.0$ ($M^* = -21.9$) mag, deriving a power law index $\alpha=-1.19\pm0.04$, and $\Phi^*_{\rm f} =11.0\pm2.0$, as plotted by the blue line in {\bf Figure \ref{fig_lfext}(a)}. 
 However, the bright part of the LF is not fit well by the single Schechter function, although the faint part is fit relatively well.
If we use a simple power law  (Log $N = - 0.4(\alpha + 1) M +$ constant, as used for Coma in  \citet{yam12}) 
for fitting the faint component (with $22.5<$F814W$\leq27.0$) only, we get $\alpha=-1.14\pm0.08$ (as plotted by the white line in the figure), which is very similar to the value from the single Schechter function fitting.

%
%


We also tried a modified form of Schecher function with a double power law form  
(with a power law index $\beta$ for the bright part and a power law index $\alpha$ for the faint part). 
This function was used to fit the double component LFs where the transition between the two components starts at $M_t$  in the previous studies \citep{lov97,pop06, bar07,ban10,mor15}:

$\Phi (L) dL = \Phi^* ( (L/L^*)^{\beta}$ exp$(-L/L^* )  [ 1+ (L/L_t)^{\alpha}] dL$ 

\noindent which is given in terms of $M$,

$\Phi (M) dM = \Phi^* 10^{-0.4(M-M^* )(\beta+1)}$ exp$(-10^{-0.4(M-M^* )}) [1 + 10^{-0.4(M - M_t) \alpha )} ] dM$.

\noindent Fitting the net LF with this function, we obtain
 $M^*_{\rm b} =-20.2\pm0.4 $, $\beta=-1.28\pm0.08$, $\alpha=-1.21\pm0.76$, $M_t =-20.7\pm0.3$ and $\Phi^*_{\rm f} =8.8\pm3.7$,
 as plotted by the red line in the figure. The value of the power law index for the faint component is also similar to those derived above.
{\bf Table \ref{tab_schechter}} lists a summary of fitting for the galaxy LFs of Abell 2744.
%
%
%

In summary, the LF of Abell 2744 shows a dip at F814W $\approx 22.3$ ($M_{r'} \approx -18.6$) mag, and the LF of the faint component of Abell 2744 is relatively flat, described with a power law index, $\alpha =  -1.14\pm0.08$.
 

\section{DISCUSSION}

\subsection{Comparison of the Galaxy LFs}

The study of the galaxy LFs has a long history (\citet{sch76,joh11,bos14, lan16,leeetal16} and references therein). The faint end of the galaxy LFs provides strong constraints to the modeling of the formation and evolution of low mass galaxies (e.g, \citet{sch88,lu15,cer16}).
However, whether the slope for the faint end of the cluster galaxy LF is steep or flat has been controversial (\citet{mil07,hars09,dep13,bos14,mar15} and references therein). 
Several studies suggested that the LFs of red dwarf galaxies in galaxy clusters at low to high redshifts are relatively flat,
with  a power law index $\alpha \sim -1.1$ to --1.4 (Virgo and Fornax \citep{jer97,hil03,rin08,lie12,dav16,gia15}, and clusters at $z\sim 0.3$ \citep{hars09}) or
with $\alpha \sim -0.8$ to --1.1 (the Hydra I and the Centaurus \citep{mis08,mis09}, and several clusters at $z<0.9$ \citep{cra09,dep13}).

On the other hand, \citet{pop06} presented a much steeper LF for SDSS galaxy clusters. They derived composite LFs of blue ($(u-r)<2.22$) and red ($2.22<(u-r)<3.0$) galaxies ($-23.0<M_{r'} <-14.0$ mag) in the SDSS galaxy clusters at low redshift. 
The blue galaxy LF is fit well by a single Schechter function with a steep faint end ($\alpha=-1.87\pm0.04$ and $M^*_r = -21.71\pm0.52$ mag). 
However, the red galaxy LF becomes flat in the bright part and rises again
at $M_{r'} \approx -18.0$ mag so that it can be described better  by a sum of double Schechter components (see their Figure 9 and Table 2). The parameters  for the fainter component  are $\alpha=-2.01\pm0.11$ and $M^*_r = -18.14\pm0.15$ mag,  and those for the brighter component are  $\alpha=-0.75\pm0.09$ and $M^*_r = -20.57\pm0.14$ mag.
Later \citet{bar07} presented, from the analysis of the galaxy LFs of nearby Abell clusters, the LFs of red galaxies show even steeper faint ends ($-5.26<\alpha<-2.83$).  They also found that the faint ends ($R_C < -18.0$ mag) of the LFs for red galaxies  
are steeper in the outer region than in the inner region of the clusters. However their data are more than one magnitude shallower than the \citet{pop06} sample so that the uncertainties for the faint ends are larger.
In addition, \citet{ban10} presented an LF of the galaxies in Abell 1689 derived from the F606W and F814W WPFC2 images, which shows an upturn 
at F814W(AB)$\approx 21$ mag. 
The faint end at $20<$F814W(AB)$<24$ mag ($19.6<$F814W(Vega)$<23.6$ mag, $-20.1<M_r' < -16.1$ mag) of the red sequence galaxies in this cluster shows a steep LF  with $\alpha=-1.97\pm0.25$ ($\alpha=-2.09\pm0.44$ for all galaxies).
 To make the story more interesting,
\citet{mil07} and \citet{yam12} presented a very steep LF of the faintest compact dwarf galaxies with $-13<M_{r'}<-10$ mag in Coma, with a logarithmic slope, $\alpha = -3.38\pm0.28$
 (derived using $log N = -0.4(\alpha +1 ) M + {\rm const}$). 
 In contrast, \citet{agu14} and \citet{agu16} found recently, using spectroscopic members of Abell 85 (at  $z=0.055$), no evidence for a steep upturn of the faint-end slope in the LF of red dwarf galaxies, obtaining 
$\alpha = -1.46^{+0.18}_{-0.17}$ for $-19.0<M_{r'}<-16.0$.

The LF of the faint end in Abell 2744 derived in this study is
relatively flat with  $\alpha = -1.14\pm0.08$, showing no evidence for faint upturn. 
\citet{dep13} used shallower $VI$ images of Abell 2744 (PI:Dupke, PID: 11689)), obtaining $\alpha = -0.95\pm0.08$
($m_I^* = 18.88\pm0.21$) for $M_I \lesssim -14.9$ mag from single Schechter function fitting (see their Fig. 11 for the CMD of Abell 2744).
Our data go much deeper than those used in \citet{dep13}, and our result for the slope of the faint end, $\alpha=-1.19\pm0.04$ derived from the single Schechter function fitting, is slightly steeper than the value given by \citet{dep13}.

The LF of the faint end for  Abell 2744  
is in strong contrast to the result for Abell 1689 given by \citet{ban10}. 
Note that the LF of galaxies in Abell 1689 presented by \citet{ban10}  shows,  in their Fig. 2(f),  an upturn at the faint end $21.0<$F814W(AB)$<24.0$ mag 
($20.6<$F814W(Vega)$<23.6$ mag), with a slope  $\alpha=-1.97\pm0.25$.
Our LF of Abell 2744 goes
much deeper ($22.0<$F814W(Vega)$<27.0$ mag) than that given by \citet{ban10}. 
%

The $B$-band LFs of faint galaxies in Virgo and Fornax given by \citet{jer97} (see also a compilation in Fig. 4 of \citet{fer12}) keep increasing until at the current completeness limits, $M_B \approx -13.2$ mag and $-13.4$  mag, respectively. 
These correspond to $M_{r'} \approx -14.7$   mag and $-14.9$ mag, respectively, for $(B-r)=1.5$. 
The LFs decrease thereafter until at  $M_{r'} \approx -12.0$ mag, which is the faintest limit in the cluster galaxy sample. 
Recently \citet{gia15} found 11 ultra-faint low surface brightness galaxies with $-13<M_{r'} <-9$ mag in a  576 arcmin$^2$ field in Virgo using the Large Binocular Telescope. These galaxies are much fainter than the previous survey limits, but the covered field is only  a small fraction of the Virgo cluster. 
Thus the detection limit of the faint galaxies in Abell 2744 in this study,
F814W$\approx 27.0$ mag ($M_{r'} \approx -13.9$ mag), is somewhat fainter than  those for wide field surveys of Virgo and Fornax.
The faint end slope for Abell 2744, $\alpha = -1.14\pm0.08$ is even slightly flatter than those for Virgo and Fornax, $\alpha \sim -1.3$ to --1.4, given by \citet{jer97,rin08,lie12,dav16,gia15}, but is similar to the value, $\alpha = -1.11\pm0.10$, for dwarf galaxies with $M_V<-9.7$ mag in Fornax given by \citet{hil03}.


The LF of the red sequence galaxies in Abell 2744
shows a dip at F814W $\approx 22.3$ ($M_{r'} \approx -18.6$) mag. The presence of this dip in the galaxy LF has been known in the low density regions of galaxy clusters, while it is rarely seen in the high density regions of galaxy clusters \citep{mer06,pop06,ban10,agu16,leeetal16}.
The dip magnitude of Abell 2744 is similar to the values in the previous studies, $M_r \approx -18\pm1$ mag.  
It has been suggested that the presence of this gap can be explained by the efficiency of merging of intermediate luminosity galaxies   depending on the galaxy density \citep{mil04,mer06,agu16,leeetal16}. Low density regions like the galaxy groups, poor galaxy clusters, or outskirt of rich galaxy clusters have a higher efficiency of galaxy merging than that of the high density regions, because they have lower velocity dispersion. Merging of intermediate luminosity galaxies (with a dip magnitude) produce brighter galaxies, while the LF of the faint galaxies changes little.  It results in a dip in the LF of galaxies, showing two components in the galaxy LF.
It is noted that the region of Abell 2744 used in this study is located at $R \ll R_{vir}$ with relatively high density of galaxies, but it shows clearly a dip. This may be related with the dynamical status of Abell 2744, which is still involved with merging of substructures.

\begin{figure}
\centering
\includegraphics[scale=0.7]{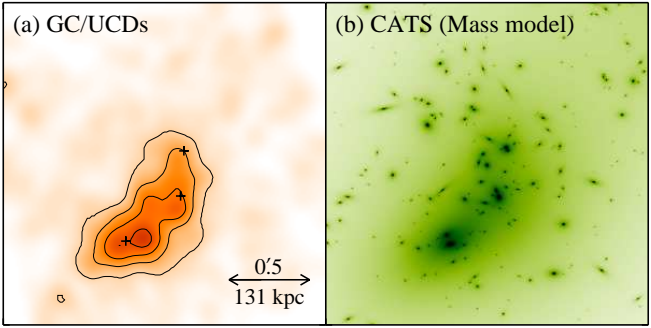} 
\caption{ Comparison of the number density maps of the GC/UCDs 
with the mass map derived using  gravitational lensing models.  
(a) All GC/UCD candidates with $0.0<(F814W-F105W)<1.0$  and $F814W<29.3$ mag.
(b) The mass density map derived with the CATS gravitational lensing models \citep{jau15}.
}
\label{fig_mass}
\end{figure}

\subsection{Comparison of Distributions of GC/UCDs and Dwarf Galaxies, and the Mass Map based on Gravitational Lensing Models}

In {\bf Figure \ref{fig_mass}} we compare the spatial distribution of the GC/UCDs 
with the mass map for the same field derived from the gravitational lensing models by the CATS team, which is provided in the HFF \citep{jau15}.  
The mass map derived from the lensing model shows a strong concentration around the two BCGs (see also \citet{wan15}).
Thus  the spatial distribution of the GCs and UCDs is similar to the mass map.
However, the degree of central concentration of the GCs and UCDs is much stronger than that of the mass density, as shown below.

\begin{figure}
\centering
\includegraphics[scale=0.9]{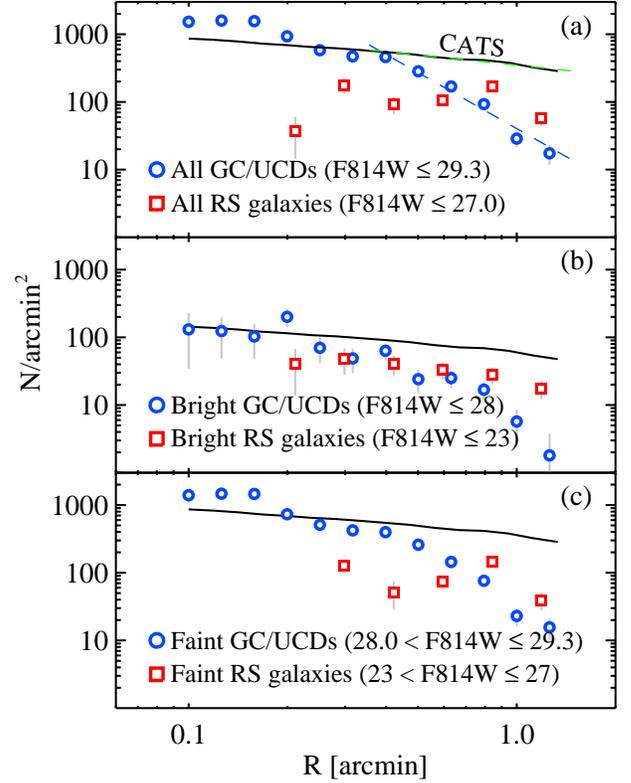}
\caption{Comparison of the radial number density profiles of the GC/UCDs (circles), those of the red sequence dwarf galaxies (squares) in Abell 2744 derived in this study, and the radial mass density profile of the CATS gravitational lensing model of Abell 2744 \citep{jau15} (solid line): 
(a) for the entire samples; 
(b) for the bright samples (the UCDs with F814W$<28.0$ mag, and the red sequence galaxies with F814W$<23.0$ mag); and
(c) for the faint samples (the UCDs with $28.0<$F814W$<29.3$ mag, and the red sequence galaxies with $23.0<$F814W$\leq27.0$ mag).
Two dashed lines in (a) indicate the power-low fit results for the profiles of the mass density map (green line) and the GC/UCDs (blue line) for the outer region at $R>0\farcm35$. 
}
\label{fig_raddencomp}
\end{figure}

For comparison with the radial number density profiles of the UCDs and GCs, we derived the radial number density profiles of the red sequence galaxies in Abell 2744, plotting them in  {\bf Figure \ref{fig_raddencomp}}. 
We subtracted the background contribution using the HXDF data. 
We divided the red sequence galaxy sample into two groups according to their magnitude:
a bright group (bright galaxies with F814W$\leq23.0$ ($M_{r'} \leq -17.9$) mag),
and a faint group (dwarf galaxies with $23.0<$F814W$\leq27.0$  ($-17.9<M_{r'} \leq -13.9$) mag).
The number of galaxies detected in the innermost region $R<0\farcm2$ ( 52 kpc) is too small for analysis so that only the number density profiles of the galaxies at $R>0\farcm2$ are useful for analysis. 
In addition, we derived the radial mass density profile of Abell 2744 from the CATS model image \citep{jau15} provided in the HFF. 
As the center for deriving the radial profiles, we adopted the center of CN-2 as in the case of the GC/UCDs.

{\bf Figure \ref{fig_raddencomp}} shows a few distinguishable features as follows. 
First, 
the radial number density profiles of the red sequence galaxies are almost flat in the outer region at $0\farcm35<R<1\farcm0$, while the profile of the bright group decreases slowly as the clustercentric distance increases.
It is noted that the radial number density profile of the GC/UCDs is much steeper than the profile of the red sequence galaxies. 
Second, 
the mass density radial profile of Abell 2744 derived from the CATS gravitational lensing models decreases slowly as the clustercentric distance increases. The slope of the mass density profile is much flatter than that of the radial number density profiles of the GC/UCDs.
The logarithmic slope of the mass density profile for the outer region at  $R>0\farcm35$ is derived from the power fits to be
$d\log \sigma / d\log R =-0.5$. On the other hand,
we derive $d \log \sigma {\rm (GC/UCD)} / d\log R = -2.75\pm0.16$ from the number density profile of the GC/UCDs, which is $\sim$6 times steeper than that of the mass density profile.
However, the mass density profile is consistent with that of the radial number density profiles of the red sequence galaxies in the outer region at $R>0\farcm35$.
It is noted that, although the spatial distribution of the GC/UCDs shows a strong correlation with the mass density map (see {\bf Figure \ref{fig_mass}}), the GC/UCDs show a significant difference in the radial distribution in the sense that the GC/UCDs show a much stronger central concentration.
These results show that the radial distribution of the GC/UCDs in Abell 2744 is much steeper than the distribution of dark matter, which is consistent with the case of
Abell 1689 \citep{ala13}.
The radial distribution of the dwarf galaxies follows well that of dark matter.

%
%

%
%

\subsection{The Origin of the UCDs}

Since the discovery of UCDs, the origin of the UCDs has been controversial \citep{hil11,nor11,nor14,bek15,zha15,liu15,jan16}.
There are three types of scenarios suggested to explain the origin of the UCDs.
First, the UCDs are just the massive end of the normal GCs \citep{mur09,chi11,ren15}. 
Second, they are the nuclei of dwarf galaxies that were stripped due to the tidal interaction with their host galaxies \citep{bek03,pfe13,pfe14,pfe16,jan16}. Direct evidence for this scenario has been found recently by detection of the existence of supermassive black hole in some UCDs \citep{set14}, 
and by finding that some UCDs show an  extended star formation history (\citet{nor15}, Ko et al. (2016, in preparation)). 
Metal-rich UCDs with young populations may be the remnants of a dissipative merger of two gas-rich dwarf galaxies \citep{bek15}.
Third, the UCDs are the remnants of primordial compact galaxies that have formed early around their host galaxies  \citep{dri04}.
This was suggested to explain the fact that the spatial distribution of the UCDs in the Fornax cluster show a stronger central concentration than than of the dE galaxies.
The origin of the UCDs may be a combination of these, rather than a single scenario \citep{nor11,nor14,pfe16,jan16}.

Recently \citet{zha15} and \citet{liu15} studied the properties of 97 UCDs (with $10<r_{\rm eff} < 100$ pc, $18.5<g<20.5$ mag, $-12.6<M_g<-10.6$ mag) in M87, a cD galaxy in Virgo, This is the largest sample of UCDs in a single galaxy. They found that there are significant differences in the radial number density profiles and the kinematics between the UCDs and the GCs in Virgo and that the mean color of the UCDs is consistent with that of the blue GCs. From this they  concluded that the origin of most UCDs in Virgo is close to the nuclei of stripped dwarf galaxies, rather than to the massive end of the GCs. This conclusion is in strong contrast to the result for the UCDs in Coma that \citet{chi11} concluded, that most UCDs in Coma are of the star cluster-origin, noting the similarity in the properties and spatial distribution of UCDs and GCs. Their sample of UCDs covers a similar magnitude range, $-13<M_V<-11$ mag.

In the discussion of compact stellar systems (with $6<r_e < 500$ pc and $2 \times 10^6 < M/M_\odot < 6 \times 10^9$) in various galaxies and galaxy clusters,  \citet{nor11} and \citet{nor14} pointed out that there are two types of UCDs: star cluster-type UCDs
and galaxy-type UCDs. The upper magnitude limit of star cluster type UCDs is $M_V \approx -13.0$ mag (or $7 \times 10^7~M_\odot$) . They noted that the LF of the UCDs and compact galaxies shows a break at $M_V  \approx -13.0$ mag. It is almost flat at the bright range ($M_V < -13.0$ mag), but it increases suddenly at the faint end ($M_V \geq -13.0$ mag) (see Figure 12 in \citet{nor14}).
The bright component represents galaxy-origin UCDs, while the faint one does star cluster-type UCDs. However, this is based on the sample made of heterogeneous data and the number of the sources brighter than $M_V = -13.0$ mag is only a handful,
as described by  \citet{nor14}, so that it is only indicative, requiring further confirmation.

The LFs of GC/UCDs in Abell 2744 in this study ({\bf Figure \ref{fig_lfpoint}}) show a break at F814W $\approx 28.0$ mag, which corresponds to $M_{r'}  \approx -12.9$ ($M_V  \approx -12.4$) mag. 
Thus this value is 0.6 magnitude fainter than 
$M_V \approx -13.0$ mag,  noted for the UCDs in the local universe \citep{nor11,nor14}. Our result based on the large sample of homogeneous data for Abell 2744 confirms the presence of the break, but at a 0.6 magnitude fainter value, $M_V \approx -12.4$ mag. 
In addition, it is noted that the mean color of the faint dwarf galaxies with
$26<F814W<27$ mag in Abell 2744 is similar to that of the UCDs/GCs.

In summary, there are
three clues useful to understand the origin of the UCDs in Abell 2744 found in this study:
(a) the presence of a break at $M_r \approx -12.9$ mag in the LF of the UCD/GCs,
(b) the radial number density profiles of the UCDs  that are much steeper than 
those of the red sequence galaxies,  and 
(c) the mean color of the faint red sequence galaxies that is similar to that of the UCDs.
These clues support the hypothesis for the dual origin of the UCDs: 
bright UCDs are the nuclei of dwarf galaxies that were stripped when they passed the inner region of a galaxy cluster, while faint UCDs are the upper end of mass functions of the GCs.

\section{SUMMARY AND CONCLUSION}

We detected a large population of GCs, UCDs, and dwarf galaxies in the deep high resolution F814W and F105W images of the sourthern core of Abell 2744 in the HFF. We could estimate the background field contribution using photometry of the parallel field and the HXDF. Then we investigated
the photometric properties of these sources in comparison with the results for the parallel field and the HXDF. To date, Abell 2744 at the redshift of $z=0.308$ is the most distant system where GCs and UCDs are studied in detail. Primary results are summarized as follows.

\begin{itemize}

\item
CMDs of the extended sources derived from deep F814W (restframe $r$) and  F105W (restframe $I$) images of Abell 2744
show a red sequence reaching down to $M_{r'} \approx -13.9$ mag. 

\item
CMDs of the point sources in Abell 2744
show a rich population of point sources whose colors are similar to those of typical GCs. 
They are as bright as
$26.0<$F814W$<29.5$ ($-14.9<M_{r'}<-11.4$) mag, and their color distribution shows a peak at (F814W-F105W)$\approx 0.5$.
They are mostly bright GCs and UCDs in Abell 2744.

\item
Th spatial distribution of the GCs and UCDs  shows a strong central concentration to the brightest galaxies. This is consistent with the mass map based on  gravitational lensing analysis \citep{jau15,wan15}.
However, the radial number density profile of the GCs and UCDs is much steeper than  the radial mass density profile based on gravitational lensing models \citep{jau15,wan15}. The mass density profile is consistent with the radial number density profile of the dwarf galaxies in Abell 2744. 

\item 
The LF of the GCs and UCDs in Abell 2744 shows a break at  F814W$\approx 28.0$ ($M_{r'} \approx -12.9$) mag. This is consistent with the one  based on the heterogeneous data for the  UCDs in the local universe \citep{nor11,nor14}. 

\item 
The number of UCDs with F814W$<28.6$ mag in Abell 2744 is estimated to be $N=147\pm26$.  
The number of GCs is derived to be $N_{GC} = 385,044\pm24,016$,  This value is 
even larger than that for Abell 1689, $N_{GC}=162,850^{+75,450}_{-51,310}$, given by \citet{ala13}. Thus Abell 2744 hosts the largest number of GCs among the known systems.

\item 
The LF of the red sequence galaxies for a large range of magnitudes $18.0<$F814W$<27.0$ ($-22.9<M_{r'} <-13.9$) mag shows double components, with a dip at F814W $\approx 22.3$ ($M_{r'} \approx -18.6$) mag.
The  faint end of the red galaxy LF is fit well by a flat power law with $\alpha=-1.14\pm0.08$, showing no faint upturn.  

\item These results support the hypothesis for the dual origin of the UCDs: 
bright UCDs are nuclei of dwarf galaxies that were stripped when they passed the inner region of a galaxy cluster, while faint UCDs are massive GCs at the upper end in the mass function of the GCs. 

\end{itemize}

The authors are grateful to John Blakeslee for providing their results on Abell 2744 presented at the AAS meeting,
and to Brian Cho 
for his help in improving the English in the draft.
This work was supported by the National Research Foundation of Korea (NRF) grant
funded by the Korea Government (MSIP) (No. 2013R1A2A2A05005120).
This work based on observations obtained with the NASA/ESA Hubble Space Telescope, retrieved from the Mikulski Archive for Space Telescopes (MAST) at the Space Telescope Science Institute (STScI). STScI is operated by the Association of Universities for Research in Astronomy, Inc. under NASA contract NAS 5-26555. 
This work also utilizes gravitational lensing models produced by PIs Bradac, Natarajan \& Kneib (CATS), Merten \& Zitrin, Sharon, and Williams, and the GLAFIC and Diego groups. This lens modeling was partially funded by the HST Frontier Fields program conducted by STScI. STScI is operated by the Association of Universities for Research in Astronomy, Inc. under NASA contract NAS 5-26555. The lens models were obtained from the Mikulski Archive for Space Telescopes (MAST). 



\end{document}